\documentclass[aps, prb, twocolumn, floatfix]{revtex4}

\usepackage{graphics}
\usepackage{epsfig}
\usepackage{dcolumn} 
\usepackage{subfigure}
\usepackage{float}
\usepackage{morefloats}
\usepackage{flafter}
\usepackage{afterpage}

\newcommand{\fig}{./}

\newcommand{\Phy }{$\Phi$ }
 
\newcommand{\cd }{${\it j_i }$ }
\newcommand{\cdnosp}{${\it j_i}$ }

\newcommand{\localexp }{$\beta_{local}$ \hspace{1pt}}
\newcommand{\localexpnosp }{$\beta_{local}$}

\newcommand{\goodgap}{
    \hspace{\subfigtopskip}
    \hspace{\subfigbottomskip}
    \vspace{\subfigtopskip}
    \vspace{\subfigbottomskip}
}

\newcommand{\threshvolt }{$V_T$ \hspace{0.75pt}}

\newcommand{\csigma }{$C_{\Sigma }$ \hspace{0.75pt}}
\newcommand{\csigmanosp }{$C_{\Sigma }$}
\newcommand{\cinter }{$C_{I}$ \hspace{0.75pt}}
\newcommand{\cinternosp }{$C_{I}$}
\newcommand{\esigma }{$E_{c_{\Sigma }}$\hspace{0.05pt}}
\newcommand{\esigmanosp }{$E_{c_{\Sigma}}$}
\newcommand{\eovercsigma}{$\frac{e}{C_{\Sigma}}$}

\newcommand{\crossvolt }{{$\Delta V_{X}$} \hspace{0.75pt}}

\newcommand{\mtr}{$\overline{\Gamma}$}
\newcommand{\ltth }{L$^{\frac{2}{3}} $~}
\newcommand{\ltthnosp}{L$^{\frac{2}{3}}$}

\newcommand{\lmotw }{L$^{-\frac{1}{2}} $~}

\newcommand{\lotwnosp }{L$^{\frac{1}{2}}$}
\newcommand{\loth }{L$^{\frac{1}{3}} $~}
\newcommand{\lothnosp }{L$^{\frac{1}{3}} $}
\newcommand{\lfth }{L$^{\frac{5}{3}} $~}
\newcommand{\leth }{L$^{\frac{8}{3}} $~}
\newcommand{\lfthnosp }{L$^{\frac{5}{3}}$}
\newcommand{\ftth }{$\frac{2}{3}$}
\newcommand{\ftthwsp }{$\frac{2}{3} $~}
\newcommand{\fttw }{$\frac{3}{2}$}

\newcommand{\ffth }{$\frac{5}{3}$}

\newcommand{\feth }{$\frac{8}{3}$}

\newcommand{\fotw }{$\frac{1}{2}$}

\begin{document}
\title{Effects of disorder on electron transport
       in \\ arrays of quantum dots}

\author{Shantenu Jha}  \author{A. Alan Middleton} 
\affiliation{Department of Physics, Syracuse University, Syracuse, 
             NY, 13244, USA} \date{\today}

\begin{abstract}
  Using analytical and numerical methods, we investigate the
  zero-temperature transport of electrons in a model of quantum dot
  arrays with a disordered background potential, where the electrons
  incoherently tunnel between the dots. One effect of the disorder is
  that conduction through the array is possible only for voltages
  across the array that exceed a critical voltage $V_T$. We
  investigate the behavior of arrays in three voltage regimes: below
  the critical voltage, above, but arbitrarily close to, the critical
  voltage, and further above the critical voltage. For voltages less
  than $V_T$, we find that the features of the invasion of charge onto
  the array depend on whether the dots have uniform or varying
  capacitances. We compute the first conduction path at voltages just
  above $V_T$ using a transfer-matrix style algorithm.  Though only
  the first path can be studied using this technique, it can be used
  to elucidate the important energy and length scales. We find that
  the geometrical structure of the first conducting path is
  essentially unaffected by the addition of capacitive or tunneling
  resistance disorder. We also investigate the effects of this added
  disorder to transport further above the threshold.  We find that
  qualitative behavior is dominated by the presence of the background
  potential, rather than capacitive or tunneling disorder, at least as
  long as these additional disorders do not have an extremely broad
  distribution.  We use finite size scaling analysis to explore the
  nonlinear current-voltage relationship near $V_T$. The scaling of
  the current $I$ near $V_T$, $I\sim(V-V_T)^{\beta}$, gives similar
  values for the effective exponent $\beta$ for all varieties of
  tunneling and capacitive disorder, when the current is computed for
  voltages within a few percent of threshold. We do note that the
  value of $\beta$ near the transition is not
  converged at this distance from threshold and difficulties in
  obtaining its value in the $V\searrow V_T$ limit.
\end{abstract}

 \keywords{} \maketitle

\section{Introduction}
It is now possible to engineer arrays of nanoparticles (1-2 nm in
diameter) in various geometrical configurations~\cite{wybourne02} and
to lithographically fabricate arrays of low capacitance islands
separated by tunnel junctions with reasonable control over array
parameters~\cite{kurdak00_1}.  Such quantum dot arrays (QDA) have been
the subject of intense investigation recently~\cite{sohn,
  wingreen+sohn}.  In spite of the relatively well controlled
properties of these arrays however, there are limitations on the
homogeneity of such systems.  Disorder at the sub-micron length scales
arises due to a variety of reasons, is inevitable and significantly
influences the properties of these otherwise well ordered arrays.  For
ligand coated nanoparticles the variation in coating properties and
separation result in different resistances to electron tunneling.  As
a consequence of the poly-dispersion in the sizes of metallic
nanoparticles, the charging energies of the individual islands differ.
Similarly for lithographically fabricated tunnel junctions, islands
with variable charging energies arise due to a dispersion of island
sizes or due to fluctuating capacitative coupling between dots and the
underlying gate.  Given the pervasiveness of random background
charges, nonuniform charging energy and fluctuations in tunneling
resistance across the array, an important focus of the current work is
to study the effect of these on transport properties of electrons.
Due to limitations of the fabrication process, it is difficult to
control the different types of disorder independently, whereas this
can can be relatively easily addressed by computer simulations.

There are many dynamical systems in which strongly interacting
particles exhibit collective transport in a random
environment~\cite{dfisher98}.  Even though the underlying microscopic
details are different, systems like the vortex glass in type-II
superconductors and charge density waves~\cite{dfisher87}, share some
general features in the long wavelength limit, e.g., they are both
characterized by the presence of a well defined threshold force below
which the system is essentially static and above which the system has
a non-linear response.  Electron transport in disordered QDA also
provide a useful system to study problems of qualitative similarity.
An advantage of QDA is that the primary interactions and the
fundamental physics are relatively better understood and arguably
under greater experimental control.

Using a combination of analytic and numerical techniques, we
investigate electron transport at zero temperature in arrays of
disordered small capacitance islands which are capacitatively
uncoupled to their neighbors. We use this system both as a model for
collective transport of discrete charges in a random environment and
for better understanding the role of disorder.  By studying similar
systems, but for different values of parameters, different regimes of
the collective transport problem can be addressed.  These regimes may
be characterized by the relative strengths of disorder, tunneling
rates and electron-electron interaction. These regimes are accessible
experimentally too, as arrays can be fabricated with varying degree of
tunability of the coupling between the array elements~\cite{marcus95}.
For example, the model in Refs.~[\onlinecite{Enomoto99},
\onlinecite{Kosterlitz98}] is similar to the model we study in this
paper -- in that offset charge disorder is included, although Gaussian
distributed as opposed to uniformly distributed -- but the screening
length is assumed infinite. (Transport in this regime is believed to
belong to the same universality class of two dimensional magnetic
vortex model in disordered superconducting films.)
Ref.~[\onlinecite{hirasawa98}] which investigates 2DEG at
semiconductor heterointerfaces designed to keep the self-capacitance
and disorder low -- the I-V characteristic is better explained by
charge soliton injection as no threshold voltage is observed.  Such
systems can be used to study two dimensional Coulomb gases which
undergo charge Kosterlitz-Thouless (KT) transitions.

\subsection{Outline \label{sec:outline}}

There have been many experimental and simulation papers
~\cite{jaeger01, jaegerprl04, berk98, berk95, ancona02} (to cite just
a few) which have used the theory developed in
Ref.~[\onlinecite{mw93}] by Middleton and Wingreen (MW). This paper
expands on the MW discussion of electron transport in disordered
arrays.  The original model and its extension to include other forms
of disorder are described in the remainder of this section.
One-dimensional arrays -- both below and above threshold -- are
discussed in detail in section~\ref{sec:onedim}.  In
section~\ref{sec:subthresh} results for 2D arrays below the threshold
voltage are presented including several results not discussed
previously.  As the original MW paper sketched only briefly the
connection between the independent conducting paths and the properties
of a directed polymer in random media (DPRM), a major aim of
section~\ref{sec:paths} and~\ref{sec:subthresh} is to establish the
connection on a more rigorous basis. In section~\ref{sec:paths} we
discuss the morphology and current carrying properties of the first
conducting path at threshold for QDA. It also provides some of the
details required to understand the non-linear scaling of current (I)
with voltage (V), which is addressed in section~\ref{sec:2d_dynamics}.

There have been several papers that have used numerical approaches to
investigate transport in arrays (both 1D and 2D) in the presence of
random background charges as well as other types of
disorder~\cite{likharev_prb03, melsen97, johansson, ancona02}.  Some
approaches, have used discrete event simulation techniques to model
the individual tunneling events, whereas some have explicitly used
computed transition rates in a master-equation approach.  The common
aim is to compute the general I-V characteristics, which as a
consequence of the collective behavior of electron tunneling is
non-trivially dependent on the individual rates.  We focus on a
statistical physics approach to the problem, thus laying a theoretical
basis for the scaling exponents observed experimentally and
numerically.

\subsection{The  Model \label{sec:model}}
The three main energy scales of QD are the charging energy
(\esigmanosp), the electron in-a-box energy levels ($\Delta$) and the
thermal energy ($kT$).  As a consequence of the small size of these
islands and tunnel junctions the capacitance involved are in the femto
to atto-Farad range, thus the charging energy -- which is the increase
in energy due to the addition of a single electron is given by $e^2/
C_{\Sigma}$ -- of these islands is large.  A characteristic feature of
QD is the clear separation of internal energy scales $\Delta$ and
\esigma.  An external energy scale ($kT$) is set by the temperature of
interest, which determines the levels that are resolved and
participate in transport.  When \esigma $\gg$ $kT$ the role of thermal
fluctuations can be ignored.  Depending upon the temperatures of
interest, $\Delta$ maybe comparable to $kT$ or different; for $kT \gg
\Delta$, the discrete energy level spectrum of the QD do not play a
role during transport.  Metallic dots are different from
semi-conducting dots by the fact that typically the level spacings for
metallic dots are much smaller compared to other energies.  At
sufficiently low temperatures, the scale of which is set by, \esigma
$\gg$ $kT$, the addition of a single extra electron to an dot
increases the dot energy; in spite of the increased energy, the dot is
stable to thermal energy fluctuations, which in turn makes it
unfavorable for more electrons to tunnel onto the same dot, resulting
in its blocking other electrons onto the dot.  This is called the {\it
  Coulomb blockade} regime.

The parameters required to characterize QDA can vary over a large
range of values and consequently so do the properties of QDA.  Thus,
it is instructive to understand the parameter space of QDA in order to
appreciate the details of the model.  The main parameters used to
characterize QDA, as opposed to individual quantum dots (QD) are: the
tunneling resistance ($R_T$) which to a first approximation is a
measure of how well confined the electrons are on a dot, the inter-dot
capacitance (\cinternosp) and the dot capacitance (\csigmanosp) which
is a function of the junction, gate and self-capacitances.  The
relative values of \csigma and \cinter are important as it determines
the extent of electrostatic coupling between dots in the array.  The
exact value of \csigma depends on the system under consideration.  For
example, typically the self-capacitance of lithographically prepared
arrays is negligible compared to the other capacitances, thus \csigma
is a function of the dot-gate and tunnel junction capacitances (for
example in Ref.~[\onlinecite{berk98}] \csigma = C$_g$ + 4C). For
nanoparticles with diameters of a few {\it nm}, the self-capacitance
becomes important and should possibly be considered in the computation
of \csigmanosp \cite{wybourne01}. Either way, \csigma still sets the
scale for the charging energy.  Independent of the actual experimental
setup considered, as long as \csigma $\gg$ \cinternosp, the dots are
considered to be capacitatively uncoupled to each other and the
electrostatic energy is determined by on site interactions only.
However if \csigma is comparable or less than \cinter the dots are
capacitatively coupled. A screening length ($\lambda$) can be thought
of the distance (in units of dots) upto which the charge on a dot can
be felt electrostatically, i.e., distance that an excess charge placed
on a dot will effect neighboring dots by polarization.  The
polarization decreases exponentially with $\lambda$, which in turn
decreases with the ratio of $\frac{C_I}{C_{\Sigma}}$; this is
consistent with understanding that there is a stronger screening of
the electrons on a dot from electrons on adjacent dots as the
capacitative coupling between a dot and the back gate increases.  For
\csigma $\gg$ \cinter, $\lambda \approx
{(\frac{C_I}{C_{\Sigma}})}^{1/2}$.

The main modification to the original MW model is the introduction of
nonuniform dot capacitance \csigma and tunneling resistances $R_T$.
The effects of underlying charge impurities trapped at the interfaces
and in the substrate are captured in a random background charge on
each dot. The effect of background charges is modeled as offset
charges on each dot ($q_i$).  The offset charge at any site is
considered to be [0,1[, as any value outside this range will be
compensated by electron hopping. Arrays with only offset charge
disorder are referred to as UC (uniform capacitance) systems.  The
area and capacitative coupling of an island to the underlying electron
gas varies from dot-to-dot in an array.  These fluctuations in the
dot-gate and self-capacitance of dots, along with stray capacitances
are incorporated by assuming a varying dot capacitance \csigmanosp.
As \csigma controls the charging energy of the dot, a non-uniform
\csigma results in different charging energies \esigma of dots.
Arrays with both offset charge disorder and a varying \csigmanosp, are
referred to as DC (disordered capacitance) systems.  Fluctuations in
the tunneling resistance -- either due to varying distance between
metallic dots or the varying material properties of the tunnel
junction separating the metallic islands between dots -- is captured
by assuming a log-normal distribution of tunneling resistances. Arrays
that incorporate variation in tunneling resistance as well as offset
charge disorder, but with a fixed value of \csigmanosp, are referred
to as RT (resistance disorder) systems.

We assume small metallic islands are separated from each other by
tunnel junctions of resistance $R_T$ but capacitatively coupled to
neighboring dots (\cinternosp).  We assume a constant capacitance
\cinter between neighboring dots and between the left and right leads
and dots adjacent to them.  The dots are assumed to be separated from
an underlying back gate by an insulating layer.  Each dot is
capacitatively coupled to the back gate with a capacitance
\csigmanosp. The leads and back gate are assumed to have infinite self
capacitance. As a consequence of the proximity of the back gate to the
dots \csigma $\gg$ \cinternosp, the screening-length is taken to be
less than one lattice spacing. Consequently the capacitative coupling
between dots is neglected.

We will consider arrays where the single-energy levels of the dots are
essentially a continuum at the Fermi level in the strongly Coulomb
blockaded (\esigma $\gg kT$) regime.  Thus tunneling is between levels
determined by \esigma.  Where a spread in values of $R_T$ is
considered, we assume tunneling resistance between any two dots is
still sufficiently large to consider electrons localized on a site
($R_T \gg h/{e^2}$). This is the regime of the ``orthodox theory'' of
single electron tunneling and is applicable for both the micron sized
lithographically defined SET (e.g., metal islands embedded in a
substrate \cite{berk95} and separated by tunnel junctions or
semi-conductor islands separated by barriers \cite{marcus95}) as well
as the 3D metallic grains.  According to the ``orthodox
theory''~\cite{orthodox} of a tunneling event across a tunnel
junction, tunneling rates (transition probability per unit time)
associated with an event are given by,
\begin{equation}
\label{eq:rate}
\Gamma = {\Delta E \over e^2R_T}{ 1 \over [1-\exp(-\frac{\Delta E}{k T})]}
\end{equation}

where $\Delta E $ is the difference in the free energy of the system
before and after the tunneling event, $R_T$ is the tunneling
resistance of the junction involved in the tunneling event, T the
temperature and $k$ is the usual Boltzmann constant. The kinetic
energy gained by the tunneling electron is assumed to be dissipated.
The value of $R_T$ is assumed to be much greater than $h/e^2$.  This
essentially implies that the wavefunction of electrons are localized
to a single dot which permits the number of electrons on any single
dot to be treated as a classical variable.  It should be pointed out
that the orthodox theory is still valid for arrays in the limit $C_i$
$\gg$ \csigmanosp, but not for dots in the other limits of $R_T \ll
R_Q$ and \esigma $\ll kT$.

In this limit the energy is all electrostatic and is determined by a
capacitance matrix $C_{ij}$ and is represented as:
\begin{equation}
  \label{eq:energyexpression}
   E   =   V_L  Q_L   +  V_R   Q_R   +  \frac{1}{2}  \sum_{ij}{(Q_{i}   +
q_{i})}{C^{-ij}}{(Q_{j}+q_{j})},
\end{equation}
where $Q_L$ ($Q_R$) are the charges on the left (right) leads, which
are at voltages $V_L$ ($V_R$) and $C^{-ij}$ is the inverse of the
matrix of capacitances between dots $i$ and $j$.  The diagonal
elements of $C_{ij}$ are the sum of all capacitances associated with a
dot and the off-diagonal elements are the negative of the inter dot
capacitances. Hence for a N$\times$N array in the limit of
$\frac{C_I}{C_{\Sigma}}$ $\rightarrow$ 0, the capacitance matrix is a
N$\times$N diagonal matrix. 

In the limit of small screening length (less than 1 dot spacing) and
the presence of offset charge disorder the voltage on dot i is given
by $V_i$ is $(Q_i + q_i)/C_{\Sigma} $.

At zero temperatures the expression~(\ref{eq:rate}) for tunneling
rates reduces to
\begin{equation}
\label{eq:zeroT_rate}
\Gamma = {\Delta E \over e^2R_T}{\Theta(\Delta E)}
\end{equation}
hence a charge may tunnel from dot i to j, only if such an event
lowers the free energy of the array i.e.
\begin{equation}
\label{eq:condition}
 V_i > V_j + e/C_{\Sigma}
\end{equation}



\section{1D arrays \label{sec:onedim}}

Before attempting to understand the detailed properties of two
dimensional arrays, we begin by an attempt to understand the
relatively simpler case of a linear chain of quantum dots, as they
facilitate an understanding of some of the ideas required later.
There have been several experiments aimed at understanding the
conduction properties of essentially one dimensional arrays of
nanoparticles \cite{tinkham99, wybourne01}.  The ability of metallic
nanoparticles to be patterned using polymers templates makes them
attractive candidates for potential future self-assembling electronic
devices.

\subsection{Uniform \csigmanosp: Insulating State   
\label{subsec:onedim_uniform}}

\begin{figure}
  \subfigure[]{\epsfig{figure= \fig 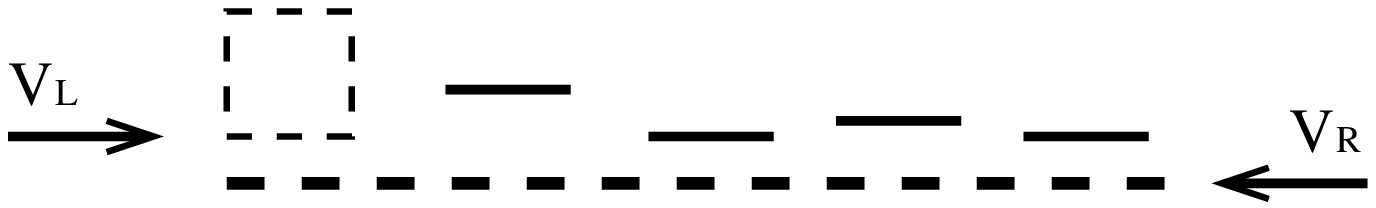, scale=0.5} 
       \label{firstincr}}
  \subfigure[]{\epsfig{figure= \fig 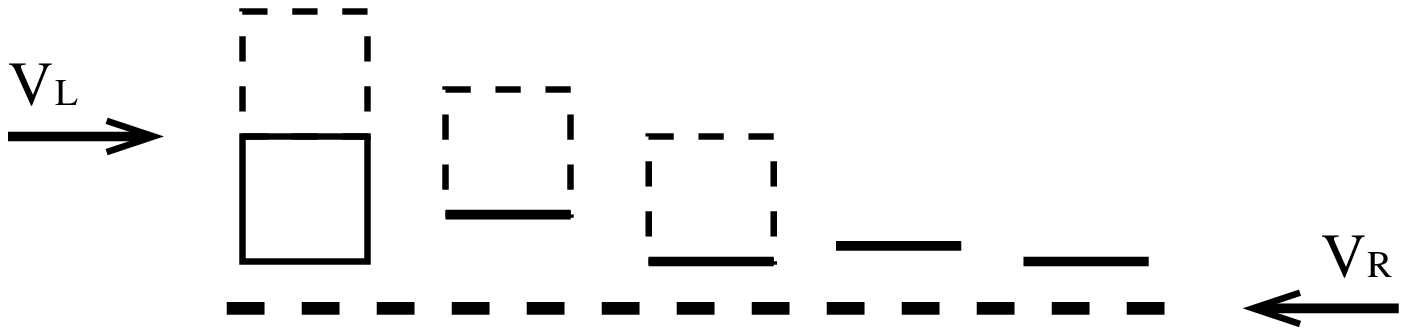, scale=0.5}
    \label{secondincr}}
  \subfigure[]{\epsfig{figure= \fig 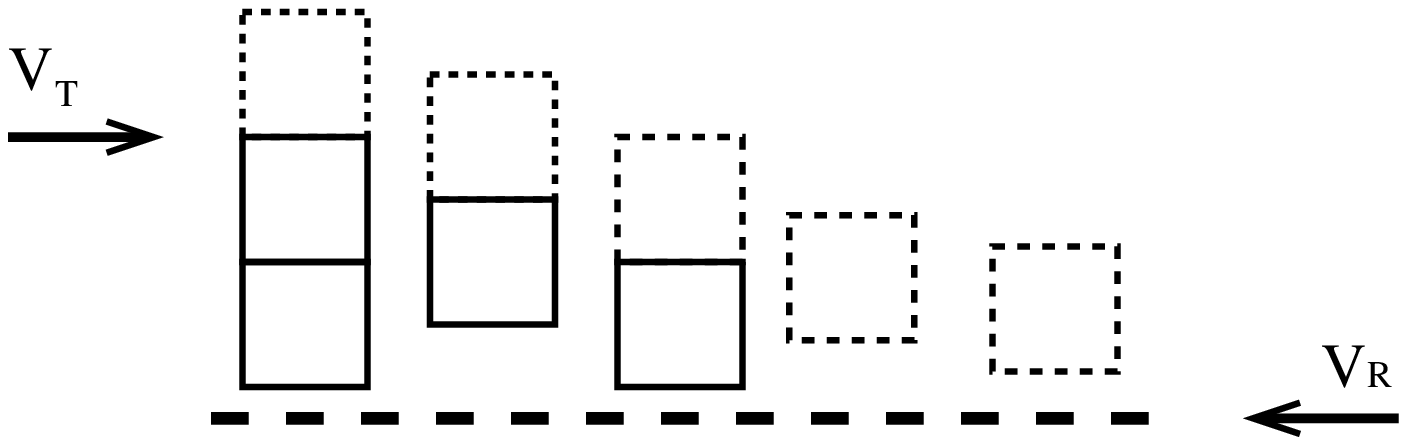, scale=0.5}
    \label{thirdincr}}
  \caption{A schematic illustrating the build of charges in a 1D array
    as the emitter voltage is progressively increased to threshold.}
\end{figure}

We start by exploring the tunneling of electrons onto the array from
the emitter lead. In the zero temperature limit, as the capacitance of
the leads is assumed to be infinite, electrons can flow onto the array
when the voltage of the emitter lead ($V_L$) is equal to or greater
than the voltage of the leftmost dot as given by
Eqn.~\ref{eq:condition}.  At this applied voltage, an electron cannot
tunnel from the leftmost dot to the next dot, say represented by the
index $i$, if dot $i$ has an offset charge impurity $q_i$ greater than
the offset charge impurity $q$ of the leftmost dot.  Electrons tunnel
onto the array only if it is possible to do so without an increase in
the free energy of the system. This is no longer possible for the
configuration in Fig.~\ref{firstincr}. In this configuration the
electron residing on the leftmost dot is considered to be {\it
  pinned}.

As shown in Figs.~\ref{secondincr} and~\ref{thirdincr} the emitter
lead voltage has to be increased by at least one unit in order that
the electrons can overcome the barrier.  As the value of the emitter
lead voltage is successively increased, there will be a cascade of
electrons tunneling onto the array from the lead, until they get
progressively pinned and it is no longer energetically possible for
electrons to penetrate further into the array. The flow of charges
onto the array at a given emitter lead voltage until they are all
pinned due to the disorder and thus no further electrons can tunnel
onto the array constitutes an {\it avalanche}.

\begin{figure}
  \subfigure[]{\epsfig{figure= \fig 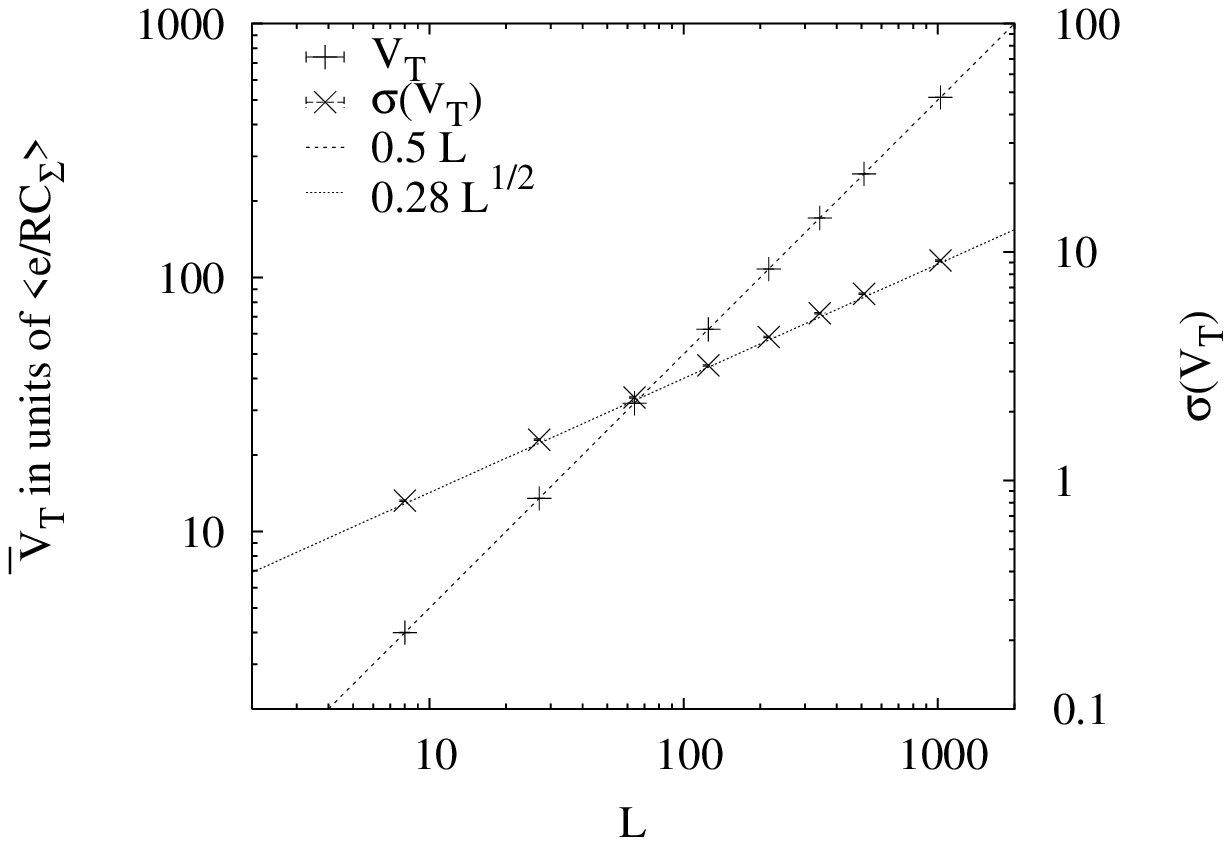, scale=0.6} 
    \label{1D.vthresh.uniform}}
  \subfigure[]{\epsfig{figure= \fig 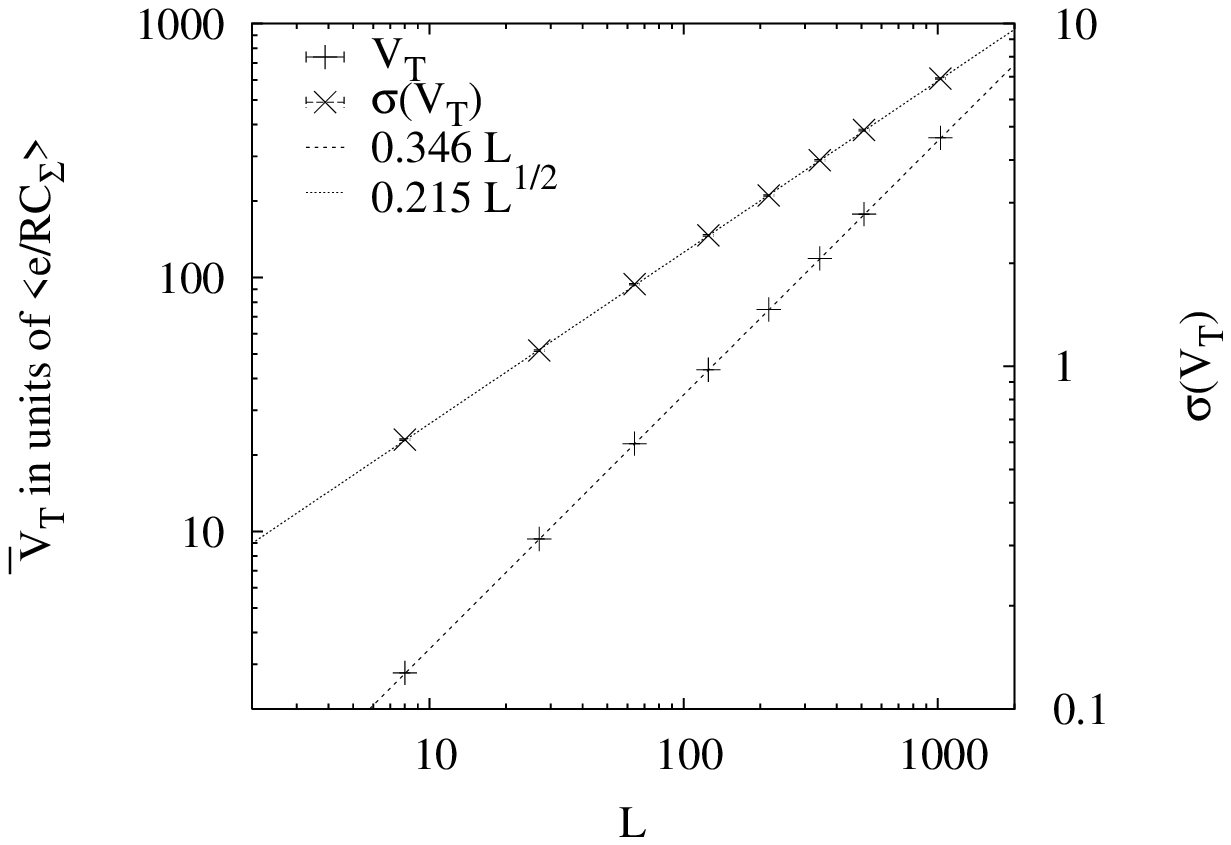, scale=0.6}
    \label{1D.vthresh.nonuniform}}
  \caption{In Fig.\ref{1D.vthresh.uniform} a plot of the scaling of $V_T$ and
    root mean square fluctuations of $V_T$ with system size for a 1D
    array when the \csigma is uniform. In
    Fig.\ref{1D.vthresh.nonuniform} scaling of $V_T$ and root mean
    square fluctuations in $V_T$ with system size for a 1D array but
    with non-uniform distribution of \csigmanosp.}
\end{figure}

There exists a unique value of the emitter lead voltage -- which
depends upon the underlying disorder profile -- at which electrons
will be able to reach the collector lead for the first time
(Fig.~\ref{thirdincr}).  This well-defined voltage value is referred
to as the threshold voltage ($V_T$). \threshvolt separates the
conducting phase from an insulating phase.  Typically in order to
reach the collector lead end of an array L dots long, an electron will
have to overcome $\frac{L}{2}$ upward steps.  These steps can be can
be understood as the average number of steps a random walk in 1D makes
in a given direction, thus the mean threshold voltage should be given
by
\begin{equation}
\label{eq:threshold}
 {\overline{V_T}} = (\frac{L}{2})(\frac{e}{C_{\Sigma}})
\end{equation}
where the over-bar represents an averaging over disorder realizations.
Sample-to-sample fluctuations in the $V_T$ can be thought of as the
root-mean-square fluctuations of a random walk in 1D which scales as
$N^{1/2}$, where N is the number of steps of the random walk. Hence
fluctuations in $V_T$ should scale with system size as
\begin{equation}
\label{eq:threshold_fluct}
 {\sigma(V_T)} \sim L^{1/2}
\end{equation}
The scaling of both $V_T$ and $\sigma(V_T)$ with system size, as shown
in Fig.~\ref{1D.vthresh.uniform} are consistent with the above
explanation.

\subsection{Uniform \csigmanosp: Conducting State}
The threshold voltage represents the lowest voltage at which electrons
can tunnel across the array, hence for $V_L > V_T$, current flows
through the array. For a given disorder realization, \threshvolt
depends on the number of up-steps encountered due to the offset charge
impurities.

If $V_L$ is marginally greater than $V_T$, so that $\nu \equiv (V_L -
V_T)/V_T \ll 1$, then the discreteness of charge and offset charge
impurities play a crucial role in determining the current. At these
voltages the current is determined by the slowest tunneling rate
($\Gamma_{slow}$) between any two neighboring dots in the array
(analogous to the net flow of traffic being determined by the
bottleneck in the path of flow), which on the average is given by
$\frac{V_L - V_T}{eRL}$, where L is the number of dots in the 1D
array.

This can be understood as follows: $(V_L - V_T)$ represents the
voltage increment over $V_T$. In principle the voltage drop can be
anywhere between between 0 and $V_L - V_T$ for a given pair of dots.
For an array with L dots there are L+1 ($\sim$ L when L is large)
voltage drops (tunneling rates), hence the minimum voltage drop across
any two dots is on the {\it average} $\frac{V_L - V_T}{L}$, which
results in a tunneling rate given by Eqn.~(\ref{eq:zeroT_rate}) to be
$\frac{V_L - V_T}{eRL}$.  Using $V_T = \frac{eL}{2C_{\Sigma}}$ we get
$\Gamma_{slow}$ = $\frac{V_L - V_T}{2RC_{\Sigma} V_T}$.  As $I =
e\Gamma_{slow}$ we have
\begin{equation}
\label{eq:current_1D}
I = (\frac{e}{2RC_{\Sigma}})\nu.
\end{equation}

\begin{figure}
  \subfigure[]{\epsfig{figure=\fig 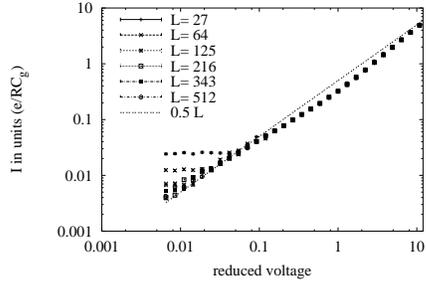,
      scale=0.45}\label{1D_curr_unif.eps}}
  \subfigure[]{\epsfig{figure=\fig 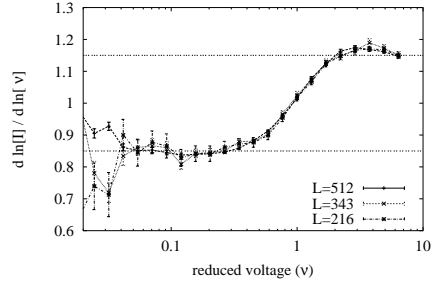,
      scale=0.45}\label{eff_exp.1.0.eps}}
  \subfigure[]{\epsfig{figure=\fig 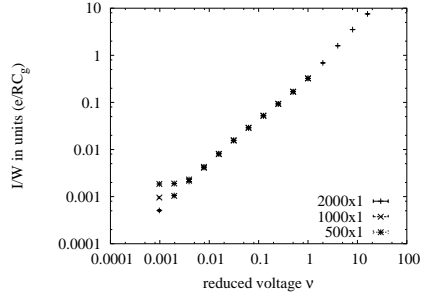, scale=0.45}
    \label{all_curr.1.0.1D.eps}}
  \subfigure[]{\epsfig{figure=\fig 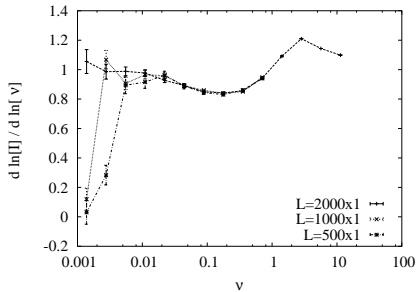, scale=0.45}
    \label{all_run_curr.1.0.1D.eps}}
  \caption{\label{curr.1D.1.0}
    I-V curves for 1D array with uniform capacitances drawn on a
    log-log plot.  Fig.~\ref{eff_exp.1.0.eps} is the plot of the local
    slope of the current versus $\nu$ derived from the data in plot
    Fig.~\ref{1D_curr_unif.eps}.  The value of the local exponent is
    computed using the values of the current at neighboring values of
    $\nu$, and assigned a value equal to the geometric mean of the two
    $\nu$. For clarity small systems have been separated from the
    largest three 1D chains simulated.  In general at low voltages the
    current scales linearly with reduced voltage ($\nu$). Flat regions
    for smaller system sizes arise because the corresponding increase
    in voltage is less than 1 unit, so although the rate to tunnel
    onto the first dot increases, the rate at the ``slow point'' does
    not change.  Consequently there is no change in the amount of
    current flowing through the array.  Theoretically the current is
    expected to scale linearly at low $\nu$ and again at high $\nu$
    with a crossover region in between. As shown in
    Fig.~\ref{eff_exp.1.0.eps}, the observed effective exponent is
    approximately 0.85 for smaller system sizes, however it is close
    to 1.0 for the largest 1D arrays at the smallest $\nu$ values as
    shown in Fig.~\ref{all_curr.1.0.1D.eps} and
    Fig.~\ref{all_run_curr.1.0.1D.eps}. }
\end{figure}

As can be seen from simulation results in
Fig.~\ref{all_run_curr.1.0.1D.eps}, in the limit of low $\nu$ and for
large system sizes, the local value of the exponent is consistent with
1. Note that the for smaller system sizes (L $\le$ 1000) the effective
exponent is quite far away from 1.  The fact that chains at least
larger than 1000 are required is an important observation.  We will
revisit its implication later in the chapter.

In the opposite regime of a high applied voltage, $\nu \gg 1$, the
current is determined by the average tunneling rate across a pair of
dots.  This is given by the $\frac{1}{eR}$ of the average voltage drop
across a pair of dots, $\frac{V_L - V_T}{L}$, i.e.,
${\overline{\Gamma}} = \frac{V_L - V_T}{eRL}$, which gives the same
scaling expression for the current with $\nu$ as
Eqn.~(\ref{eq:current_1D}).  For values of $\nu \sim 1$, a crossover
from slow point dominated current linear scaling to high applied
voltage linear scaling is observed.

\subsection{Non-Uniform \csigmanosp: Insulating State}

\begin{figure}
  \subfigure[]{\epsfig{figure= \fig 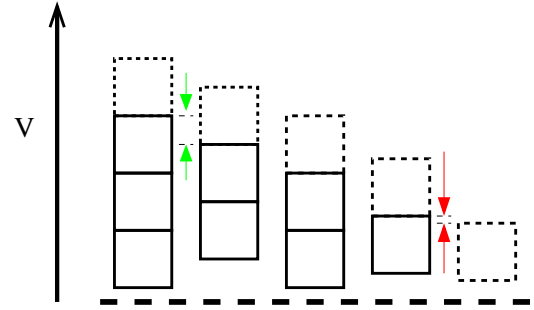, scale=0.6}
    \label{spunif_schm}}      
  \subfigure[]{\epsfig{figure= \fig 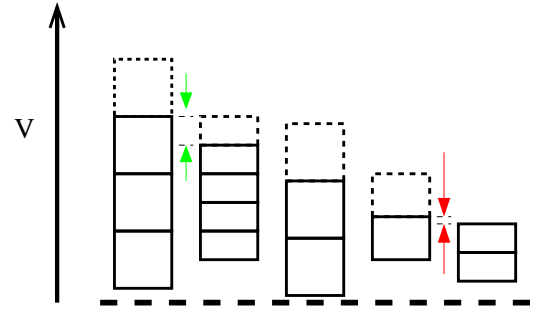, scale=0.6}
    \label{spdis_schm}}
  \caption{Fig.~\ref{spunif_schm} is a schematic of dot voltages in a 1D array
    of dots with uniform \csigmanosp. The block heights indicate the
    increase in potential on the addition of an electron. The block
    heights are the same for different dots as \csigma is the same.
    The broken line boxes at the top indicate the potential if another
    electron were added. The left arrow indicate the difference in
    potential that will determine the rate of tunneling when the
    occupation number of the leftmost dot is 4 and its right neighbor
    is 2. Similarly the right arrow indicates the rate of tunneling
    when the occupation number of the last dot is 1 and the preceeding
    dot to its left is 2.  Fig.~\ref{spdis_schm} is a similar
    schematic when \csigma is non-uniform, with \csigma for the
    second, fourth and fifth dots different.}
\end{figure}

As mentioned, the introduction of dots with non-uniform \csigma leads
to an array with dots of different charging voltages.  For our
simulations, we assume \csigma to be uniformly distributed between 1.0
and a maximum fluctuation of 2.0.  If we attribute the variation in
\csigma by a factor of 2 to fluctuations in the size of the dots, it
corresponds to a change in a variation in the linear dimension of dots
by a factor of $\sqrt2$.  The determination of the \threshvolt gets
complicated by the presence of both offset charge disorder and varying
charging energies.  This is illustrated in Fig.~\ref{spdis_schm} where
the spacings between voltages are different for different dots; this
is in contrast to dots with uniform \csigma as shown in
Fig.~\ref{spunif_schm}.  The increment in voltage required in order to
tunnel between a pair of dots is due to two independent random
variables -- $\frac{1}{C_\Sigma}$ and ${\beta}_i$, where
$1/{C_{\Sigma}}$ is the charging energy of dot i and ${\beta}_i$ is
between 0 and 1 which represents the required increment due to the
offset charges. $V_T$ can be written as the
${\Sigma}_i({\beta}_i/C_{\Sigma})$, where the summation runs over the
number of dots in the array, L.  $V_T$ can therefore be written as
$L.\langle {\beta}_i \rangle.\langle \frac{1}{C_{\Sigma}} \rangle$,
where $\langle\rangle$ represents the average values. Hence
\begin{equation}
\label{eq:vthresh.dis}
{\overline{V_T}} =  \langle 1/{C_{\Sigma}} \rangle L/2,
\end{equation}
where $\frac{1}{C_{\Sigma}}$ is $\log(2.0)$ for the assumed maximum
value of 2.0. Although $V_T$ scales as L, similar to the UC arrays,
$\sigma(V_T)$ behavior is more complicated.  An exact analytical
expression for $\sigma(V_T)$ -- which can be derived using the
expectation values of $1/{C_{\Sigma}}$ and $1/{C_{\Sigma}}^2$ gives,
\begin{equation}
\label{eq:sigma_dis_vthresh.}
\sigma(V_T) = \sqrt{\frac{L}{6} - \frac{L}{4} (\log(2))^2}.
\end{equation}
Thus, up to leading order $\sigma(V_T)$ scales as $L^{1/2}$.  This is
consistent with our results as can be seen in
Fig.~\ref{1D.vthresh.nonuniform}.

An often used technique to explore the disorder energy scale is to
study the response of the system on changing the boundary
condition~\cite{braymoore87}.  For disordered 1D QDA, we change the
boundary condition at the right lead and study the change in threshold
voltage.  We define ${\Delta}V_T({\delta}V_R)$ as the difference in
$V_T$ on changing the value of $V_R$ by ${\delta}V_R$.  Recall that
for UC arrays $V_T$ was completely determinable by the number of
up-steps in the offset charge impurities.  For the uniform capacitance
case the response is trivial: a shift in the $V_R$ by ${\delta}V_R$ --
where ${\delta}V_R$ is $\frac{e}{C_{\Sigma}}$ or a
multiple thereof -- changes 
\threshvolt by the same amount.  This is a consequence of the response
being periodic in voltage. For the 1D QDA with disordered
capacitances, however, a shift in $V_R$ by ${\delta}V_R$ guarantees a
change in $V_T$ by ${\delta}V_R$ {\it only on the average}, due to a
consequence of the threshold voltage being invariant to the zero-level
of the lead voltages.  Specific values of ${\Delta}V_T$ depend upon
the specific disorder configuration.  As a consequence of the
invariance just mentioned,
\begin{eqnarray}
\label{eq:invariance1.}
\langle {\Delta}V_T\rangle  = 
P({\Delta}V_T  \neq 0)\langle{\Delta}V_T\rangle_{{\Delta}V_T \neq0}+ \nonumber \\ P({\Delta}V_T=0)\langle{\Delta}V_T\rangle_{{\Delta}V_T=0}
\end{eqnarray}
where $P({\Delta}V_T \neq 0)$ ( $P({\Delta}V_T=0)$ ) is the
probability that the threshold voltage changes (does not change),
$\langle{\Delta}V_T\rangle_{{\Delta}V_T \neq 0}$ the average of the
non-zero values of ${\Delta}V_T=0$. 
Also $\langle{\Delta}V_T\rangle_{{\Delta}V_T=0}$ is 0.

\begin{figure}
  \subfigure[]{\epsfig{figure=\fig
      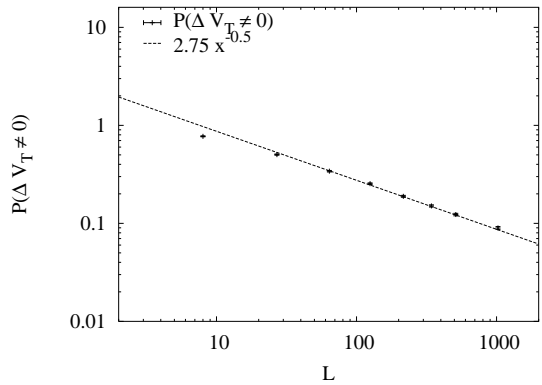,scale=0.6} \label{prob_nonzeroVt}}
  \subfigure[]{\epsfig{figure=\fig 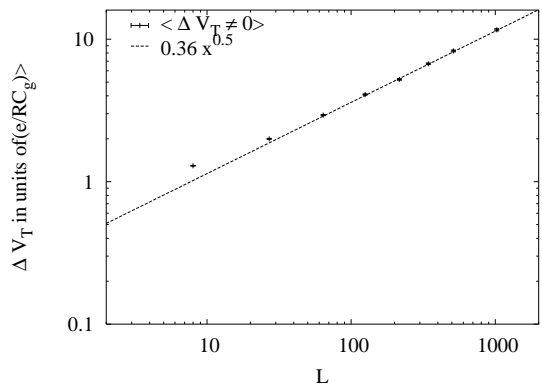,
      scale=0.6} \label{mean_nonzeroVt}}
  \caption{
    Fig.~\ref{prob_nonzeroVt} plots the probability for 1D systems
    with non-uniform \csigma that that \threshvolt changes when the
    right lead voltage is increased by one unit.  The probability
    decreases as L$^{-\frac{1}{2}}$.  For 1D systems with non-uniform
    \csigmanosp, the mean value of non-zero ${\Delta}V_T$ scales as
    the square root of the system size.  As a consequence of
    invariance, the mean value of all ${\Delta}V_T$ is equal to the
    value by which the right lead voltage is incremented.}
\end{figure}
The response to a change in the right lead voltage can be formulated
in terms of a 1D random walk problem.  Given the initial and final
points of a random walk, one can ask what is the probability that a
random walk starting a distance {\it a} from the initial point of the
original walk, intercepts the original walk before a distance $l$?  If
we assume that interception with the original walk results in
annihilation, we can ask of the surviving walks -- what is the typical
separation of the end-point of surviving walks from the end-point of
the original random walk?  It is known \cite{boltzmann_award}, that
the probability of ``survival'' decreases as $l^{1/2}$ and the typical
separation scales as $l^{1/2}$ (the square root of the mean standard
deviation of a $l$ step random walk).  The mapping to the random walk
problem is carried out by considering $V_R$ as the origin of the walk,
the potential of each dot at threshold (minus the gradient) to be the
positions of the original random walk and finally ${\delta}V_R$ as the
distance $a$ of the initial point of the second random walk from the
original random walk.  Thus, we expect that the probability of
${\Delta}V_T \neq 0$ (i.e.,
survival) 
and the mean of the non-zero ${\Delta}V_T$
($\langle{\Delta}V_T\rangle_{{\Delta}V_T \neq 0}$) should scale as 
\lmotw and \lotwnosp respectively.
This is consistent with our numerical results as shown in
Fig.~\ref{prob_nonzeroVt} , although there are significant deviations
at smaller system sizes.

\subsection{Non-uniform \csigmanosp: Conducting State}

We discussed earlier how for UC arrays in the regime of low $\nu$, the
value of the current is determined by the presence of dynamically
important slow points. An important distinction that arises in DC
arrays is that the location and value of slow points is less regular.
For UC arrays, the value of smallest voltage drop -- and hence the
minimal tunneling rate -- was bound to increase every time the emitter
lead voltage was incremented by one unit (\eovercsigma).
\begin{figure}
  \center \includegraphics[scale=0.6]{\fig 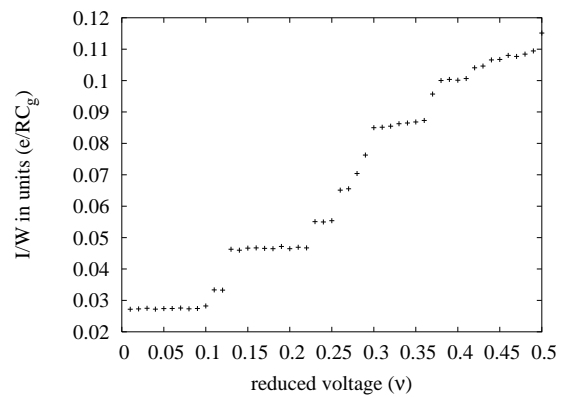}
\caption{\label{plateau}
  I-V characteristic for a single sample 1D QDA with disordered
  \csigmanosp.  Unlike UC arrays, at small reduced voltages an the
  current value remains constant over a range of $\nu$ values -- hence
  the observed plateau(s).  This happens when the smallest tunneling
  rate does not change in spite of increasing $\nu$.  When the rate at
  the slow point does change, however, the value of the current jumps
  resulting in the step like features.}
\end{figure}
Unlike UC arrays, the amount by which the emitter lead voltage must be
increased in order to overcome the slow point does not have a
well-defined lower bound and varies significantly from
sample-to-sample and with the value of reduced voltage.  As can be
seen from the presence of plateaus in Fig.~\ref{plateau}, the I-V at
low $\nu$ for a single DC array is qualitatively different to a
sample-averaged I-V.

At a given value of $\nu$, the mean tunneling rate (\mtr) is
proportional to the average potential gradient, and is given by,
\begin{equation}
\label{eq:relate_rate_nu}
\overline{\Gamma} = \frac{\nu}{2}{\langle \frac{e}{C_{\Sigma}} \rangle}.
\end{equation}
Based upon the relative values of the \mtr ~and the typical maximum
fluctuations from \mtr, we can categorize the applied voltage into
three regimes.  These regimes are: (i) when the maximum fluctuation
are larger then the mean tunneling rate; (ii) when the maximum
fluctuations are of the same value as the average gradient, and (iii)
when the maximum fluctuations are much less then the average gradient.

In regime (iii) the fluctuations about the mean gradient can be
ignored; they no longer influence the current value and consequently
the current is given by the Eqn.~(\ref{eq:current_1D}).  The average
potential profile at any given dot can be computed using the average
potential gradient and the distance of the dot from the boundary. By
subtracting the dot potentials from the averaged potential profile at
each site, we can calculate the fluctuations, and thus the roughness
of the voltage surface.  In regime (i) the roughness of the voltage
surface scales as \lotwnosp.  Assuming Gaussian distribution of the
fluctuations, the mean value of the maximum of N variables from a
distribution with mean $\mu$ and standard deviation $\sigma$ is given
by $\mu$ + $\sigma$$\sqrt{a \log N}$~\cite{kinnison}.  The deviation
from the mean tunneling rate is maximum at the slow point.  Given that
the slow point can occur anywhere between any two dots (i.e., 0 and
L), the typical value of the maximum fluctuation scales as $\mu$ +
$\sigma \sqrt{a \log (\frac{L}{2})}$. The increment in voltage
$\Delta{\nu}$ should thus be greater than the maximum barrier
(fluctuation) in order that the slow point be overcome, i.e., an {\it
  additional} electron flows over the slow point which in turn will
result in an increase in current.  Therefore the probability that a
change by ${\Delta{\nu}}$ will overcome a slow point is given by the
probability that the typical maximum fluctuation is less than
${\Delta{\nu}}$.  As ${\Delta{\nu}} \sim L$, P(${\Delta{\nu}} >$
typical maximum) $\sim \frac{L}{L^{1/2} \sqrt{a \log(L)}}$, which
approaches 1 as L gets larger.

\begin{figure}
  \subfigure[]{\epsfig{figure=\fig 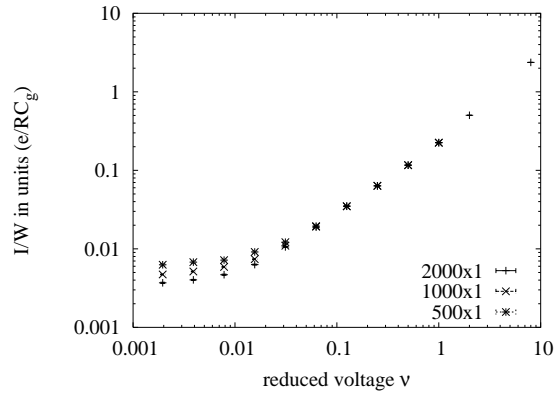, scale=0.6}
    \label{all_curr_1D.2.0}}
  \subfigure[]{\epsfig{figure=\fig 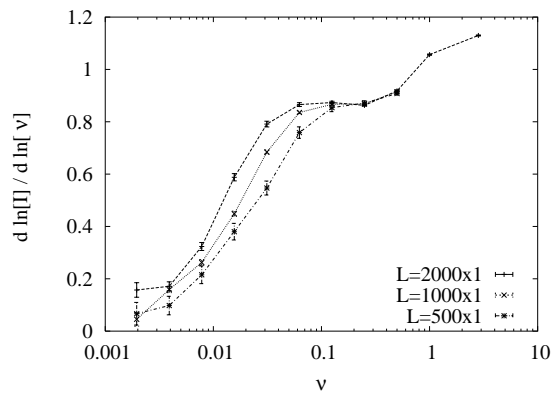, scale=0.6}
    \label{all_run_curr_1D.2.0}}
\caption{
  Analogous to Fig.~\ref{curr.1D.1.0}, the I-V curves and \localexp
  for 1D arrays with disordered \csigma.  The major difference is that
  the value of the effective exponent, even for the largest 1D arrays,
  does not appear to plateau at 1.0, but seem to flatten out at 0.85.}
\end{figure}

If step like features persist in the I-V for single samples, then
given the sample-to-sample fluctuations in the location of the
plateaus, the sample averaged I-V curve will be more or less flat. As
seen in Fig.~\ref{all_curr_1D.2.0}, there is a voltage upto which the
averaged current is more or less static. The value of this voltage
decreases with increasing system size -- consistent with the arguments
that the same increase in $\nu$ is more likely to result in a slow
point being overcome as L gets larger.  In spite of the irregular
change in the value of the minimum rate, the average value of the
minimum voltage drop across any two dots remains $\frac{V_L -
  V_T}{L}$. Hence the average value of the minimum rate remains as
before -- $\frac{V_L - V_T}{eRL}$, which implies that once the
``static current regime'' is overcome the current should scale
linearly with voltage.  Thus in spite of the introduction of variable
\csigmanosp, current scales linearly with $\nu$ in regimes (i) and
(iii), similar to UC arrays. The crossover from linear scaling in
regime (i) to (iii) -- corresponds to regime (ii) and is more
complicated to understand analytically.

In Fig.~\ref{all_curr_1D.2.0} and Fig.~\ref{all_run_curr_1D.2.0} we
plot the current and local exponent values for the largest systems (L
$\geq$ 500).  As shown in the plot of effective exponents
(Fig.~\ref{all_curr_1D.2.0}), numerically we find $\beta$= 0.85 $\pm$
0.02 in the low voltage regime (0.05 $< \nu <$0.5) and 1.2 $\pm 0.05$
in the high voltage regime ($\nu >$ 1.0).  Similar exponent values are
found for system sizes less than L=500 (not shown).

\section{2D Arrays: Insulating State}\label{sec:subthresh}

We saw in the previous section, how for one-dimensional arrays, charge
flowed onto the array from the emitter lead till it was energetically
favorable.  In this section we will attempt to develop an
understanding of the progressive build up of charge in two-dimensional
arrays, as the emitter lead voltage ($V_L$) is increased; the
tunneling of charge is still governed by Eqn.~\ref{eq:condition}. The
flow of charge onto the array can thus be viewed as lowering the
energy.  Such {\it relaxation} of charges so as to lower the system
energy, is analogous to several different systems where the system
reaches a lower energy via a series of avalanches~\cite{dfisher98,
  plastic2}.

The threshold voltage is the minimal emitter lead voltage possible
such that when electrons tunneling onto the array from the emitter
lead have sufficient potential to overcome the disorder barriers and
reach the collector lead. Given a disorder configuration it is not
trivial to determine the {\it minimal} voltage for 2D arrays.  A naive
approach might be to think of the L$ \times $W array as W, 1D arrays
of L dots each; trivially compute the ``threshold'' voltage for each
of the 1D arrays and then find the minimum.  The computed minimum
would still probably be overestimating the true threshold voltage.
Determining the threshold voltage can be formulated as an optimization
problem, but motivated by the aim of understanding the physical
buildup of charge in QDA, we take a different approach.  For a given
emitter voltage, we add charges till a meta-stable insulating state is
reached; then the emitter lead voltage is progressively increased,
building up charges until an insulating state no longer exists. The
value of the emitter lead voltage at which electrons first tunnel onto
the collector lead is our computed threshold voltage.

\subsection{Avalanches}
As briefly mentioned earlier that an avalanche at a given voltage is
the flow of charge onto the array until the flow is arrested by
disorder.  Avalanches in QDA are qualitatively similar to those found
in other systems with collective elastic transport. Some well studied
examples are vortex flow in disordered superconductors~\cite{plastic4}
and the avalanches when an interface like a CDW moves in quenched
disordered systems~\cite{narayan94, aam93}. For 2D arrays, the
location where charges tunnel in a given avalanche, helps develop the
notion of connected elastic domains~-- {\it basins}.

We define $q_{i}(V_{L}^{-})$ as the charge of site $i$ before the
emitter lead voltage is incremented to $V_L$ and $q_{i}(V_{L}^{+})$ as
the charge of site $i$ after the emitter lead voltage has been
incremented to $V_L$.  The physical size $A$ of an avalanche is the
number of sites where $q_{i}(V_{L}^{-}) \neq q_{i}(V_{L}^{+})$, and
the volume is $\sum_{i} {q_{i}(V_{L}^{-}) - q_{i}(V_{L}^{+})}$. If we
set $n(A,V_L)$ to be the number of avalanches between size $A$ and
$A+dA$, at an emitter lead voltage of $V_L$ and define $N(A)$ =
$\int_{0}^{V_T} n(A,V) dV$ , then $N(A)$ can be thought of as the
number of such avalanches that occur in going from a $V_L$= 0 to $V_L$
= $V_T$.

\begin{figure}
  \subfigure[]{\epsfig{figure=\fig 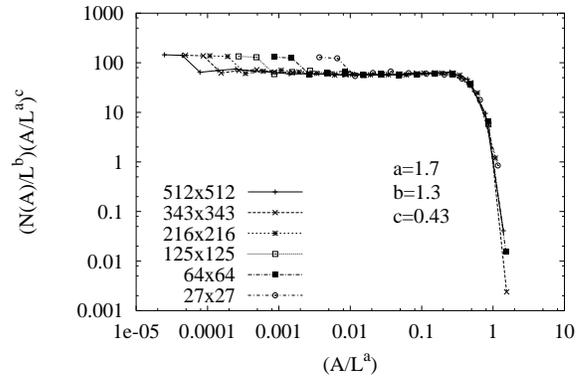, scale=0.6}
    \label{symm.NA_A.1.0.eps}} \goodgap
  \subfigure[]{\epsfig{figure=\fig 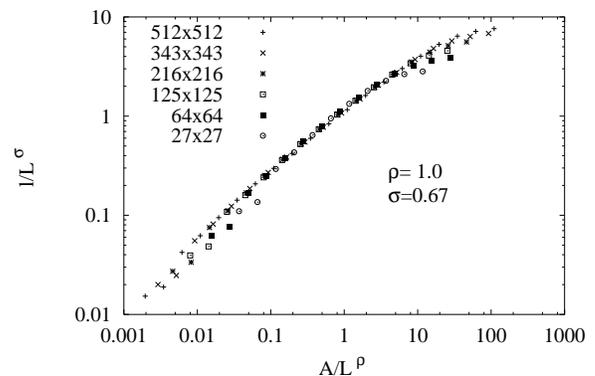, scale=0.6} 
    \label{symm.l_vs_A.1.0.eps}} \\
  \caption{
    Scaling collapse for the distribution of avalanche sizes N(A), for
    symmetrical systems with uniform \csigma is plotted in
    Fig.~\ref{symm.NA_A.1.0.eps}.  The exponent {\it a}, gives the
    typical size of the largest avalanches as L$^{a}$ ; avalanches of
    sizes greater than L$^{a}$ become increasingly improbable.
    Fig.~\ref{symm.l_vs_A.1.0.eps} shows the collapse of data for the
    mean linear size of avalanches with size between A and A+dA for
    systems with uniform \csigmanosp.  From the scaling collapse we
    estimate $d_f$ to have a value of (= $\rho/\sigma$) = 1.5.}
\end{figure}

We explore the distribution of avalanche sizes for square samples
(L$\times$L). The size of an avalanche $A$ can also be thought of as
the ``surface area'' -- which is equal to the number of dots that
electrons tunnel onto during an avalanche at a given $V_L$. As the
size of avalanches vary over several orders of magnitude -- starting
with avalanches of size 1 to system spanning avalanches -- we use
logarithmic bin sizes. Logarithmic binning is natural for exploring
power laws and reduces fluctuations in plots.

\begin{figure}
  \subfigure[]{\epsfig{figure=\fig 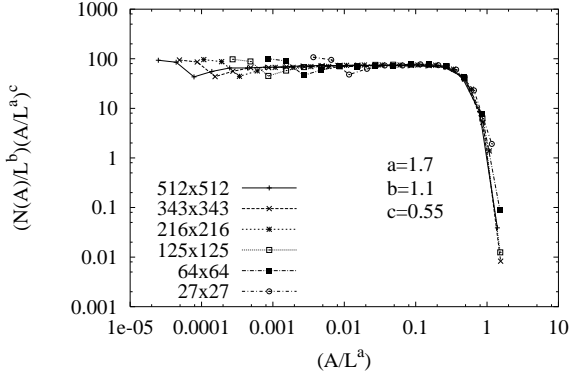, scale=0.6}
    \label{symm.NA_A.2.0.eps}} \goodgap
 \subfigure[]{\epsfig{figure=\fig 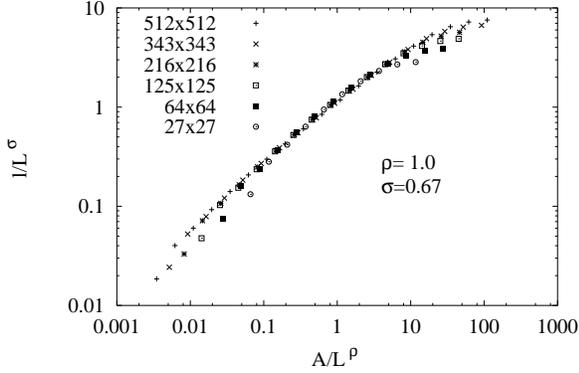, scale=0.6} 
    \label{symm.l_vs_A.2.0.eps}} \\
\caption{
  The data collapse for distribution of avalanche sizes for
  symmetrical arrays with non-uniform \csigma is plotted in
  Fig.~\ref{symm.NA_A.2.0.eps}, whilst the collapse of data for the
  mean linear size of avalanches with size between A and A+dA for
  symmetrical systems with non-uniform \csigma is plotted in
  Fig.~\ref{symm.l_vs_A.2.0.eps}. From the range of exponents for
  which the scaling collapse is acceptable we get the $d_f$ (=
  $\rho/\sigma$) = 1.5 $\pm$ for avalanches.}
\end{figure}

Using standard finite size scaling, we conjecture a scaling form for
$N(A)$ to be of the type:
\begin{equation}
  \label{eq:aval_scal_form}
  N(A) = L^{b}{\hat N}({A/L^{a}}),
\end{equation}
where ${\hat N}(x)$ scales as $x^c$ in the limit of $x \ll 1$ and
${\hat N}(x)$ approaches a constant in the limit of $x \approx 1$,
where $L$ is the length of the system.  The exponents $a$ and $b$ are
determined to be those exponents for which a scaling plot of
$N(A)/L^{b}$ vs $A/L^{a}$ yields a single scaling function ${\hat
  N}({A/L^{a}})$.  The two exponents are not independent and can be
shown to be related by the relation $a + b = 3$.~\footnote{The sum of
  the product $N(A)A$, for all avalanches upto threshold, scales as
  $L^{3.0}$ (for systems of size L$\times$L), from which the relation
  $a+b=3.0$ can be derived by using the scaling ansatz in the
  integral, $\int N(A) dA \sim L^{3.0}$. For logarithmic binning as $n
  dA = N(A) d(ln A)$, $N(A)dA = n(A)A dA$ where n(A) represents the
  number of avalanches in the linear bin [$A, A+dA$].}  In addition to
the two exponents $a$ and $b$, a third exponent $c$, can be used to
make the curve flat in the regime where $A < L^a$, which for square
systems of length and width L is related to the other two exponents by
$b - ac=2$.~\footnote{for avalanches of size $A < L^a$, N(A) scales as
  $L^{2.0}$.  Also for $A < L^a$, $N(A)/L^b \sim {(A/L^a)}^c$ where c
  is $<$ 0, which for a given A leads to $L^{b - ac} \sim L^2$, i.e.,
  $b - ac = 2.0$.}  As there are two constraints for the three scaling
exponents, we get only one independent exponent from the scaling
collapse of the distribution of the sizes of avalanches.  As shown in
Fig.~\ref{symm.NA_A.1.0.eps} the collapse of data to a single scaling
function is satisfactory, which indicates that the dimension of the
avalanches is \ffth. The typical size of the largest avalanches is
given by \lfthnosp.  To study further the morphology of the
avalanches, we compute the the mean of the maximum length of
avalanches with sizes between A and A+dA. We collapse the data as
shown in Fig.~\ref{symm.l_vs_A.1.0.eps} onto a single curve and
determine that exponents $\sigma$ and $\rho$, defined in the scaling
function:
\begin{equation}
\label{eq:linear_size_scal_form}
l(A) = L^{\sigma}{\hat L}({A/L^{\rho}}),
\end{equation}
to have values consistent with \ftth and 1 respectively.  We get a
collapse to $l/L^{\sigma} \sim ({A/L^{\rho}})^{\kappa}$ for all system
sizes L, thus $l \sim A^{\kappa}$ and the relation constraining the
exponents is therefore ${\sigma} = {\rho} {\kappa}$ or ${\kappa} =
{\sigma}/{\rho}$. We know that the $A \sim {l^{d_f}}$, where $d_f$ is
defined to be the fractal dimension, from which we get ${\kappa} =
1/{d_f}$ thus $d_f = {\rho}/{\sigma}$. From the computed values of
${\rho}$ and ${\sigma}$, $d_f$ works out to be 1.5. This is consistent
with the conclusions from the distribution of avalanche sizes.

Finally, we have also investigated the avalanche structure using the
radius of gyration~($R_g$) of avalanches, which is defined in the
usual way as:
\begin{equation}
\label{eq:rg_exp}
   {R_g}^2 = {{\sum(r_i - {\overline r})^2}\over N},
\end{equation}
and study the scaling of the mean $R_g$ for avalanches of sizes
between $A$ and $A+dA$.  Numerical evidence~\cite{jhathesis} indicates
that the scaling of the area with $R_g$ is similar to the fractal
dimension of the avalanches, which implies that the avalanche
morphology is compact, i.e., does not have any significant holes.

We have investigated avalanche structure using three ways and the
results of all three are consistent with the hypothesis that typical
avalanches are compact with dimension of \ffth.  For systems with
uniform \csigmanosp, the sequence of dots at which avalanches
originate is periodic in left lead voltage (with periodicity
\eovercsigma).  We can thus think of ``basins'' of dimension
$\frac{5}{3}$ evolving as charge flows into the array, with some
basins growing at the expense of others.  In general the basin
structure is not isotropic, as they have a preferred growth direction
and the linear size in the direction transverse to this preferred
direction grows only as $l^{\frac{2}{3}}$ where {\it l} is the linear
extent in the preferred direction.  Thus in a square samples of length
and width L, there are approximately $N_b(l) \sim L/l^{\frac{2}{3}}$
independent regions of activity, where $N_b(l)$ is defined as the
number of basins at a distance $l$ from the left lead.  Hence to
increase the chances of having the large basins that scale with L, we
simulated systems of length L and width a multiple of
L$^{\frac{2}{3}}$ (width = N$_b$ \ltthnosp).

\begin{figure}
  \subfigure[]{\epsfig{figure=\fig 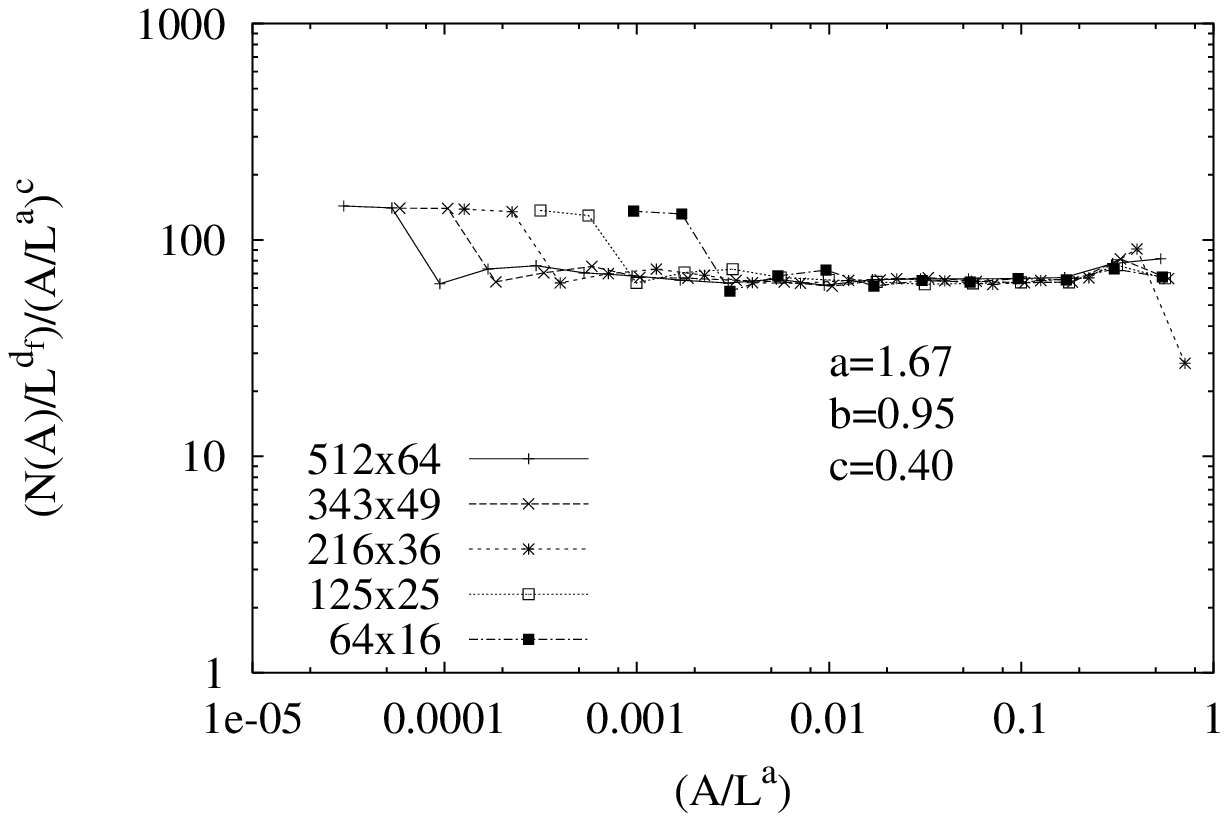, scale=0.6}
    \label{NA_A.1.0.eps}} \goodgap
  \subfigure[]{\epsfig{figure=\fig 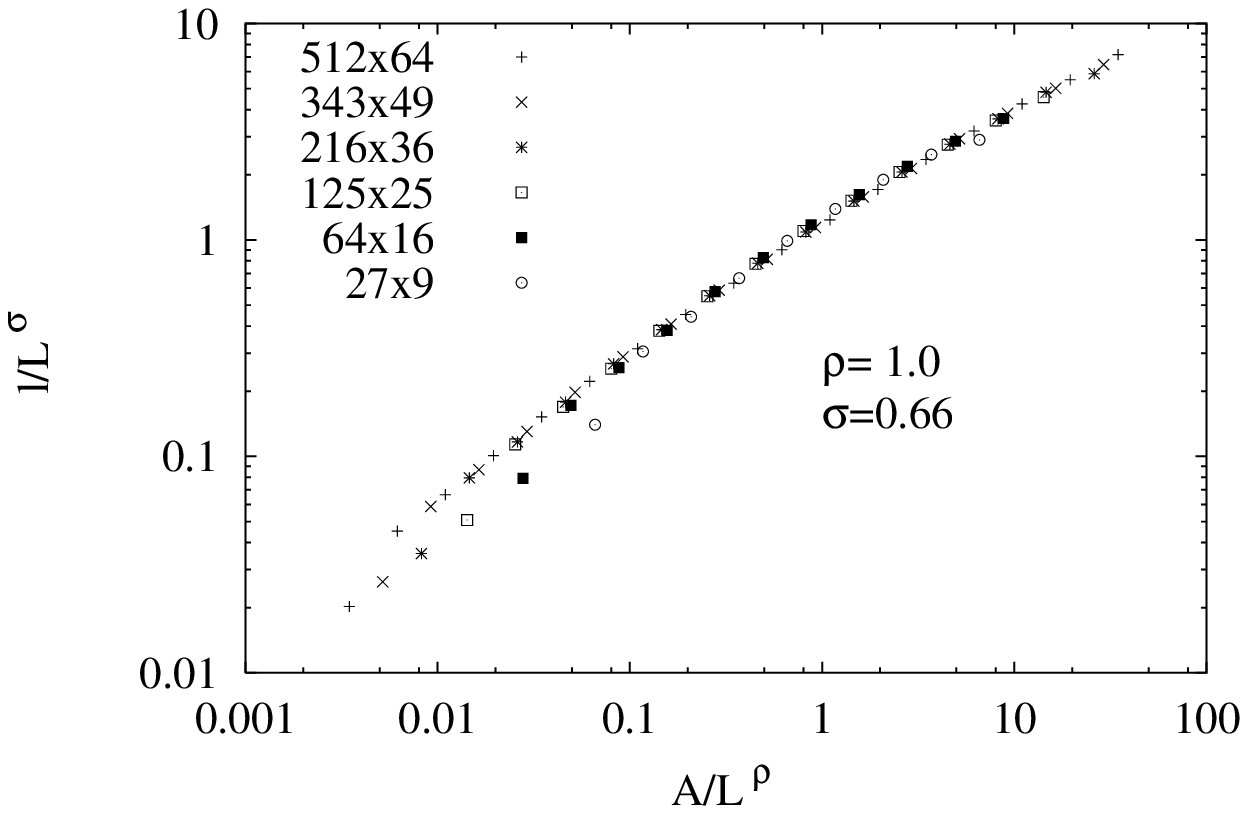, scale=0.6} 
\label{l_vs_A.1.0.eps}} \\
\caption{
  Scaling collapse for the distribution of avalanche size for UC
  arrays of size L$\times$ L$^{2/3}$. The exponent {\it a} and {\it
    c} are the same for symmetrical systems to within errors, while
  {\it b} differs. Fig.~\ref{l_vs_A.1.0.eps} plots the data collapse
  for the mean linear size of avalanches with sizes between A and A+dA
  for systems of size L $\times$L$^{2/3}$ with uniform \csigmanosp.
  Exponents are the same as those for symmetrical systems.}
\end{figure}

Similar to square samples, exponents in the scaling collapse for the
distribution of avalanche sizes for systems of size L$\times$\ltth
(Fig.~\ref{NA_A.1.0.eps}), are not independent but constrained by two
relations.

Given that the total number of dots is \lfth the sum of the product
$N(A)A$ scales as \leth rather than L$^{3.0}$; hence $a + b=$ \feth.
Also the number of avalanches in the bin [A, A+dA] scales as \lfth, so
$b - ac=$ \ffth~in this case.

An important difference in the avalanche structures between the
uniform and disordered \csigma systems is the lack of periodicity
(irregular) in emitter lead voltage and that the basins no longer
evolve by quenching other basins (they overlap).

Avalanches in DC arrays are not periodic in emitter lead voltage and
basins don't typically evolve by quenching other basins -- they tend
to overlap. This behavior is different to UC arrays.  By using the
three methods discussed earlier we find that capacitance disorder does
not affect the structure of the avalanches.
Fig.~\ref{symm.NA_A.2.0.eps} shows the scaling collapse for the
distribution of avalanche sizes N(A), for square arrays with
nonuniform \csigmanosp. The constraining equations in this case are
now $a + b = 2.8$ (as the sum of the product of N(A)A for all
avalanches $\sim L^{2.8}$) and $b - ac = 2.0$. In spite of the
presence of capacitance disorder, the value for the exponent $a$ is
the same to within errors for the value for uniform \csigmanosp.
Exponents characterizing the scaling of mean linear size and mean
$R_g$ with area for DC arrays also agree to within errors with
exponent values from UC arrays. Thus avalanches remain essentially
compact with a dimensionality of \ffth.  For L$\times$\ltth DC arrays
there is no change in the values of the exponent {\it a}, though the
constraining equations change to $a + b = 2.47$ and $b- ac =
1.67$~\cite{jhathesis}.

In this subsection, we have used finite-size analysis and been able to
successfully relate several finite-size exponents via scaling
relations.  Table~\ref{table:symbol_table} provides a quick summary of
the values and the context in which they are used.  Taken along with
the fact that these exponents and scaling relations help characterize
the transition from an insulating to a conducting state (the
conducting state is yet to be discussed), it is reasonable to view the
transition as a {\it critical transition} with associated critical
exponents and behavior.

\subsection{Interface motion}
The maximum advance of charge into the system at a given $V_L$ can be
used to define an interface. Properties of the interface can be used
to understand other properties of the system like fluctuations in
$V_T$. Some details are required about the way we define the
interface.  At a given $V_L$, there will be some dots onto which
electrons have not tunneled yet, defined relative to the original
stable configuration reached by relaxing an original configuration
with $0 < V_i < e/{C_\Sigma}$.  We refer to such dots as zero excess
dots.  We can define the interface as either of the following: (i)
contour of leftmost sites along each row which has not had an electron
tunnel onto it, or (ii) the contour of last sites along each row which
has had an electron tunnel onto it. The two although seemingly similar
are different in the sense that the second definition considers the
case where there may be ``bubbles'' of zero electron dots enclosed
behind the interface. The difference, however, is not significant as
the long wavelength properties of the interface (e.g., roughness) do
not seem to depend upon which definition is used. As $V_L$ is
increased, electrons tunnel onto arrays, via avalanches and if
electrons tunnel onto a zero excess dots, the interface advances. The
motion of the interface in response to a driving force, can be
described in the language of an elastic medium driven through a random
potential. We will argue that some quantitative correspondences exist
in fact. The dynamics of such elastic interfaces through quenched
disorder has been extensively studied in recent years \cite{barabasi},
e.g., CDW, flux lines in type II superconductors etc, fluid flow
through a porous medium to name some, flux front in thin films of type
II superconductors \cite{surdeanu99}, combustion of paper
\cite{alava1997, alava2000}.

\begin{figure}
  \subfigure[]{\epsfig{figure= \fig 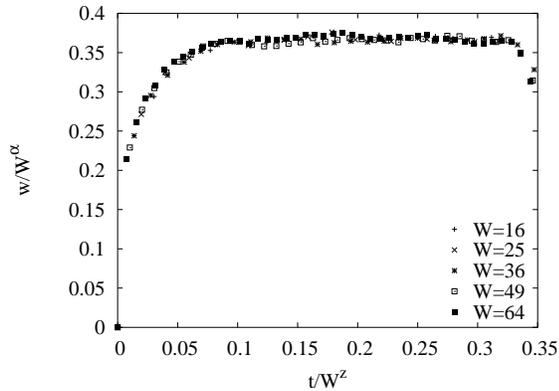,
      scale=0.6}
    \label{width_vs_time.moa=1.1.0.eps}}
  \subfigure[]{\epsfig{figure= \fig 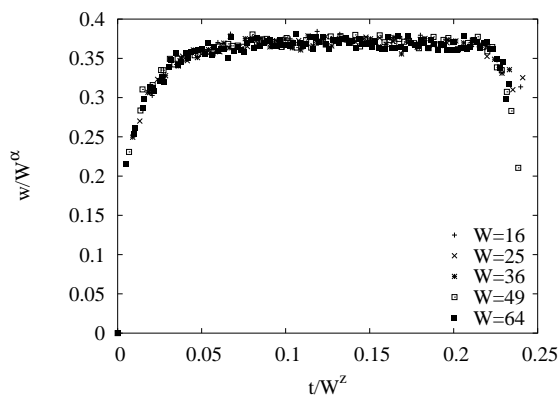,scale=0.6}
    \label{width_vs_time.moa=1.2.0.eps}}
  \caption{
    Collapse of the roughness of the interface for $L \times L^{2/3}$
    systems using the scaling form given by
    Eqn.~\ref{eq:family-vicsek}. Here W is the system width and $t$ is
    measured by the distance of the mean location of the interface
    from the emitter lead.  Fig.~\ref{width_vs_time.moa=1.1.0.eps} is
    for systems with uniform \csigmanosp, where the values of the
    exponents used in the collapse are $\alpha$= 0.5 and z=1.5.
    Fig.~\ref{width_vs_time.moa=1.2.0.eps} is for systems with
    non-uniform \csigma and the values of the exponents used in the
    collapse are $\alpha$= 0.5 and z=1.5 as well.}
\end{figure}

We define the roughness (width) of the interface as the square root of
the mean of the square of the fluctuation from the mean position. On
increasing $V_L$ the interface advances further into the system and
gets rougher.  As charge builds up behind the interface, the advance
of the interface is analogous to the growth of a surface due to
deposition of a material.  It is well known, that such surfaces become
increasingly rough with time, gradually reaching a saturation width.
For QDA as the advance of the interface is governed by $V_L$; it plays
the role of time, which upto a constant factor is the same as the mean
position of the interface.  Using the well known scaling
form~\cite{familyvicsek}:
\begin{equation}
\label{eq:family-vicsek}
w(L, t) = L^{\alpha}{\hat W}(t/L^z),
\end{equation}
we were able to collapse data on the width of the interface with time
onto a single scaling curve Fig.~\ref{width_vs_time.moa=1.1.0.eps}. We
initially used symmetrical L$\times$L systems to study the properties
of the interface. Due to the large values of the dynamic exponent z
(1.5), we were able to study only small system sizes with interfaces
with saturated width. Consequently in order to study steady state
properties of larger interface lengths, it is prudent to study
non-square systems like L$\times$\ltthnosp, thereby permitting a more
accurate determination of the exponents and hence the universality
class the interface growth process belongs to.

From the the collapse in Fig.~\ref{width_vs_time.moa=1.1.0.eps}, we
find values of the roughness exponent $\alpha$ = 0.5 and dynamic
exponent z = 1.5 -- therefore the growth exponent $\beta_g$ = 0.33.
This is consistent with the roughening of the interface being in the
KPZ universality class~\cite{barabasi}, where $\alpha$= \fotw and $z$=
\fttw.  The KPZ universality class is consistent with the symmetries
of the system, viz., rotation is a symmetry on large scales
\cite{mw93}, interactions are short range and the speed of the
interface advance lacks large fluctuations.  In light of the {\it
  assumed} lack of spatial correlation of the underlying charge
disorder for dot arrays (statistically Galilean invariant) and the
fact that the interface {\it will} move forward when the emitter lead
voltage is increased by one unit (\eovercsigma), a description of the
interface advance in terms of thermal KPZ equation seems valid.

Some avalanches involve electrons hoping onto a zero excess dot -- a
new-site.  When an avalanche involves new-sites, the interface is
reconfigured; the distribution of the avalanches that involve
new-sites provides information on the reconfiguration (advance) of the
interface.
\begin{figure}
  \subfigure[]{\epsfig{figure=\fig 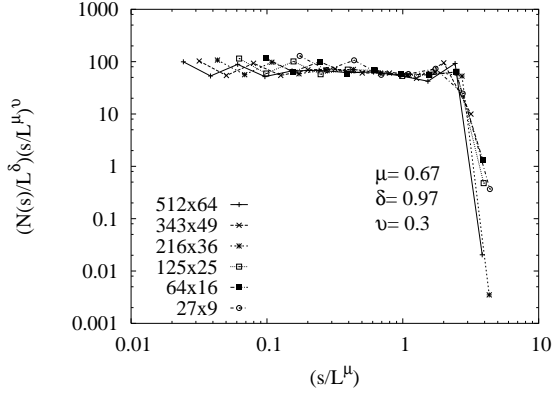, scale=0.6}
    \label{NS_S.1.0.eps}} \goodgap
  \subfigure[]{\epsfig{figure=\fig 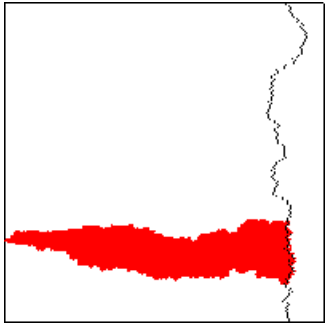, scale=0.6} 
\label{un_cross_iface.eps}} \\
\caption{
  Scaling collapse for the distribution of the number of avalanches
  involving between s and s+ds new-sites for uniform \csigmanosp. The
  exponents are related by $\mu + \delta = 1.67$ and $\delta - \mu
  \upsilon = 0.67$.  The value of $\mu = 0.67$ indicates that when the
  interface moves it typically advances by 1 dot spacing.
  Fig.~\ref{NS_S.1.0.eps} shows an avalanche crossing the interface
  for uniform \csigmanosp.  Notice that the amount by which the
  avalanche overshoots the interface is of the order one.}
\end{figure}
The scaling collapse for the distribution of avalanches that have
between s and s+ds new-sites for UC arrays is shown in
Fig.~(\ref{NS_S.1.0.eps}).  We define $N(s)$ analogous to $N(A)$,
where $s$ is the number of new-sites visited in an avalanche.  For
L$\times$\ltth samples the sum of the product $N(s)s$ for all
avalanches, scales as the number of dots in the array (\lfth). Thus
the constraint on the exponents $\mu$ and $\delta$ in the scaling
ansatz:
\begin{equation}
 \label{eq:new-site_scal_form}
 N(s) = L^{\delta}{\hat \eta}(s/L^{\mu}),
\end{equation}
is given by $\mu + \delta= 1.67$.  Another constraint is determined by
the scaling of the number distribution of avalanches with the number
of new-sites for a given bin ([s, s+ds]), with system length as \ltth,
which results in only one independent exponent in the scaling ansatz.
Hence $\delta - \mu \upsilon = 0.67$. We find that the exponent values
from the collapse consistent with these constraints. We interpret the
value of the exponent $\mu$ = 0.67 as giving the typical number of new
sites involved in an avalanche of linear length {\it l} as l$^{0.67}$.
We know that the width of the an avalanche of linear length {\it l},
is also l$^{\frac{2}{3}}$, which indicates that the avalanche
typically involves one new dot for each dot along the width.  Thus the
interface advances smoothly on the average by 1 dot along the width of
the basin of activity. Fig.~\ref{un_cross_iface.eps} shows the
configuration of the interface at a given $V_L$ and an avalanche that
crosses the interface with the portion to the right of the interface
being the new-sites involved in the avalanche. These new-sites will
determine the new configuration of the interface after the avalanche.
\begin{figure}
  \subfigure[]{\epsfig{figure=\fig 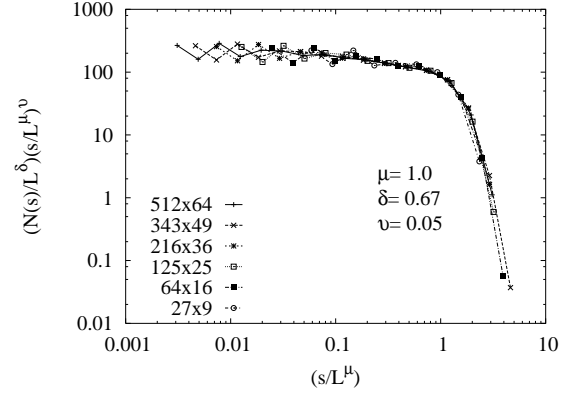, scale=0.6}
    \label{NS_S.2.0.eps}} \goodgap
  \subfigure[]{\epsfig{figure=\fig 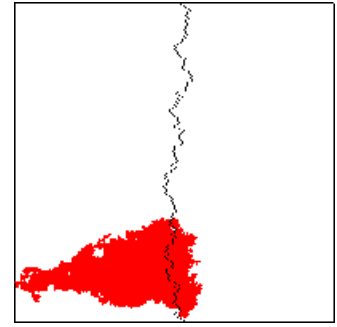, scale=0.6} 
    \label{dis_cross_iface.eps}} \\
\caption{
  Scaling collapse for the distribution of the number of avalanches
  involving between s and s+ds new-sites for L $ \times L^{2/3}$
  systems with non-uniform \csigmanosp.  The exponents are related by
  $\mu + \delta = 1.67$ and $\delta - \mu \upsilon = 0.67$.  The value
  of $\mu = 1.0$ indicates that when a segment of the interface moves
  it typically advances by $l^{1/3}$.  Fig.~\ref{dis_cross_iface.eps}
  shows an avalanche crossing the interface for non-uniform \csigma
  systems.  Typically the overshoot of the $l^{2/3}$ portion of the
  interface is $l^{1/3}$.}
\end{figure}

Further information on the movement of the interface can be obtained
by studying the voltages (V$_L$) at which an avalanche that involves
new-sites occurs, or equivalently when the interface advances.  We can
define \crossvolt as the difference in $V_L$ between two avalanches
that manage to cross the interface (there may be several avalanches
that do not cross the interface between two interface crossing
avalanches). Based upon the assumption that the advance within basins
should be independent, it can be shown~\cite{jhathesis} that
\crossvolt scales as W$/$l$^{\frac{2}{3}}$.

We've seen how the structure of the avalanches is the same
irrespective of the presence or absence of disorder in the
capacitance, even though there are changes of major significance in
the motion of the interface. If however, as shown in
Fig.~\ref{NS_S.2.0.eps} we attempt a scaling collapse for the number
of new sites covered in an avalanche the exponent values are different
from the uniform \csigma exponent values. The exponent $\mu$ has a
value 1.0 to within errors. This value can be interpreted as follows:
the dimensionality of avalanches is \ffth, which means for linear size
$l$ the width is typically $l^{\frac{2}{3}}$. When an interface
crossing avalanche occurs, the average amount by which it overshoots
the interface scales as $l^{\frac{1}{3}}$, hence covering
$l^{\frac{2}{3}}$ x $l^{\frac{1}{3}}$ new sites.

This can be seen in Fig.~\ref{dis_cross_iface.eps}, which represents a
typical interface crossing avalanche in a sample with disordered
\csigmanosp, where the avalanche overshoots the interface by a
significant amount compared to the uniform \csigmanosp (where the
overshoot was of order one spacing).  For DC arrays avalanches do not
occur with any fixed regularity -- either spatial or temporal -- so a
large number of avalanches may occur which do not reconfigure the
interface, followed by an avalanche that rearranges the interface by a
large amount. Compared to the smooth motion of the interface in arrays
with uniform \csigmanosp, the motion of the interface for DC arrays is
rather jerky.  It is important to mention that the motion appears
jerky locally, but at a coarse-grained scale and {\it on average} the
velocity of the interface is well defined and smooth till it reaches
the collector lead.

\subsection{Threshold voltages revisited}\label{sec:revisited}
Similar to one-dimensional systems the \threshvolt for two-dimensional
systems scales linearly with the system length.  For two-dimensions
however, there is an additional dependence on the width of the
samples, which can be understood using the concepts of basins and
interface advance from earlier subsections.  It also helps explain
voltage fluctuations.

\begin{figure}
  \subfigure[]{\epsfig{figure=\fig 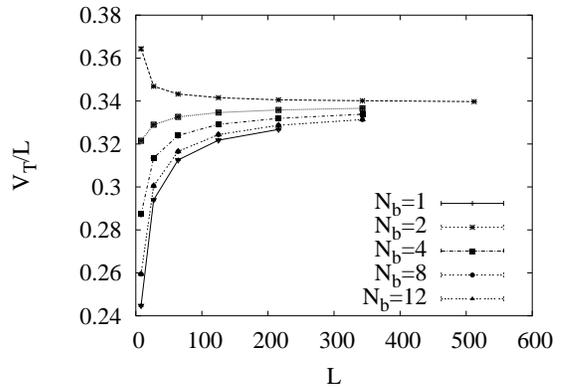, scale=0.6}
   \label{V_t-by-L.1.0.eps}} \goodgap
 \subfigure[]{\epsfig{figure=\fig 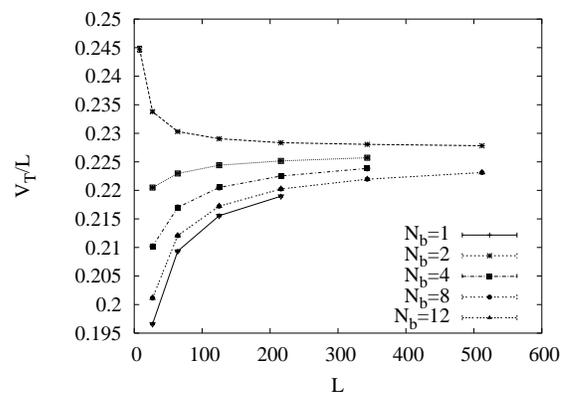, scale=0.6}
    \label{V_t-by-L.2.0.eps}} \goodgap
\caption{
  The dependence of $V_T$ on the number of basins, N$_b$(L) for
  uniform \csigma systems is plotted in Fig.~\ref{V_t-by-L.1.0.eps}.
  The dependence of $V_T$ on the number of basins (N$_b$(L)) for
  non-uniform \csigma systems is plotted in
  Fig.~\ref{V_t-by-L.2.0.eps}.}
\end{figure}
\begin{figure}
  \subfigure[]{\epsfig{figure= \fig 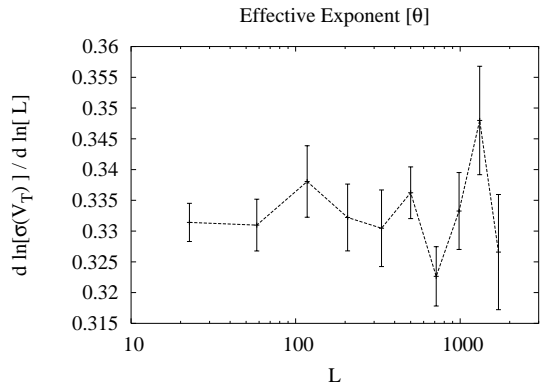,
      scale=0.6} \label{eff_exp.threshfluct.uc}}
  \subfigure[]{\epsfig{figure= \fig 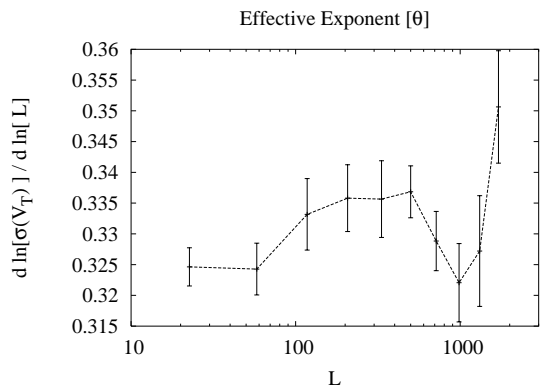,
      scale=0.6} \label{eff_exp.threshfluct.dc}}
  \caption{Effective exponents for the fluctuation of threshold voltage
    for arrays with uniform capacitance and arrays with disordered
    capacitance are plotted in Fig.~\ref{eff_exp.threshfluct.uc} and
    Fig.~\ref{eff_exp.threshfluct.dc} respectively. Sample-to-sample
    fluctuations in $V_T$ for nonuniform \csigma arrays are similar to
    the uniform \csigma and scale as $L^{0.33}$ to within errors for
    all values of $N_b(L)$.}
\end{figure}

With increasing $V_L$ the interface advances further along into the
system until finally electrons reach the collector lead at $V_T$.
Fig.~\ref{V_t-by-L.1.0.eps} shows how $V_T$ normalized by system
length (L) depends upon the width of the system studied.  When $N_b(L)
>$ 1, in addition to fluctuations within a single basin, $V_T$ is
determined by the basin that moves the interface to the right lead the
earliest.  With increasing $N_b(L)$, the expectation value of the
maximum advance of the interface at a given $V_L$ increases,
explaining the decreasing value of $\frac{V_T}{L}$.  The
sample-to-sample fluctuations in $V_T$ can be attributed to the
roughness of the advancing interface.  We saw that the roughness
exponent $\alpha$ for the interface was \fotw. Assuming a value of z=
\fttw, for L $\times$ W samples, where W = \ltth the interface reaches
its saturation width given by W$^{\frac{1}{2}}$, which is \loth. As
shown in Table \ref{table:sigmavt}, fluctuations in $V_T$ agree with
this picture.  Local values of the threshold fluctuation exponents are
plotted in Fig.~\ref{eff_exp.threshfluct.uc} and
\ref{eff_exp.threshfluct.dc} and they are consistent with the scenario
depicted.

A finite-size scaling length can be defined in terms of the
characteristic fluctuations in $V_T$. ${\sigma (V_T)}/V_T$ can be
thought of as defining the scale characterized by $L^{-1/{\nu}_T}$,
where ${\nu}_T$ is the exponent characterizing the finite-size effects
on the transition and for both uniform and disordered \csigma systems
has a value of \fttw.

Analogous to the 1D systems, we investigated the response to changed
boundary conditions -- measured by the change in right lead voltage
$V_r$ -- by measuring difference in $V_T$, as a method of probing the
disorder energy scale.  It is shown~\cite{jhathesis} that
$\langle{\Delta}V_T\rangle_{{\Delta}V_T \neq 0}$ scales as
L$^\frac{1}{3}$, i.e., for 2D when \threshvolt changes, it changes on
average by a value given by L$^{\frac{1}{3}}$.  We have discussed the
mapping between Eden growth (which is in the KPZ universality class)
and DPRM~\cite{eden}.  Using the analogy, the maximum point of advance
of the interface in our systems can be mapped to the ground state of a
DPRM.  It is known that the free energy fluctuations of the ground
states, both for sample-to-sample fluctuations and higher order
excitations scale as L$^{\frac{1}{3}}$.  For systems whose ground
state (maximum advance of interface) is unable to overcome the
increased voltage of the right lead (a change in boundary conditions)
the next lowest energy state (interface position) needs to be enough
to overcome the changed boundary condition; the energy of which is
typically L$^{\frac{1}{3}}$ higher than the ground state.  The \loth
behavior can be understood without invoking the mapping between Eden
growth and DPRM.  We saw in the subsection on interfaces, that the
mean voltage increment to move the interface so as to have a new
maximum position scaled as \loth. When the right boundary condition is
changed, either the last avalanche is able to overcome the increased
right lead voltage (in which case $\Delta V_L = 0$) or requires an
increase in $V_L$, in order to surmount the barrier at the right lead.

\begin{table}
\caption{Symbols used and comparison of values 
  for uniform and non-uniform \csigma samples.}
\begin{tabular}{cccc} 
Symbol & Used in & uniform \csigma & non-uniform \csigma\\
\hline
a, b, c                    & N(A) vs A & 1.7, 1.3, -0.43   & 1.7, 1.1, -0.55 \\  
$\rho$, $\sigma$, $\kappa$ & l vs A    & 1.0, 0.63, 0.58  & 1.0, 0.67, 0.67\\
$\alpha$, $\beta_g$, z & Family-Vicsek scaling & 0.5, 0.33, 1.5 &  0.5, 0.33, 1.5 \\
$\mu$, $\delta$, $\upsilon$ & N(s) vs s & 0.67, 1.0, -0.3 &  1.0, 0.67, -0.05\\
$\theta$   & fluctuations in $V_T$  & 0.33 & 0.33 \\
$\tau$   & mean of nonzero ${\Delta}V_T$  & 0.0 & 0.3  \\
\end{tabular}
\label{table:symbol_table}
\end{table}

\begin{table}
\caption{
Fit  values  for  $\theta$ for uniform and non-uniform \csigma samples,
with different number  of basins ($N_b$). 
For $\sigma({V_T})=  AL^{\theta}$  
both A  and $\theta$ were  fit parameters.}
\begin{tabular}{ccc}
\hline
$N_b(L)$ & uniform $C_{\Sigma}$  & non-uniform $C_{\Sigma}$  \\
\hline
1 &  0.33 $\pm$ 0.01 & 0.33  $\pm$ 0.01 \\ 
2 &  0.34 $\pm$ 0.01 & 0.34 $\pm$ 0.01 \\  
4 &  0.35 $\pm$ 0.01 & 0.34  $\pm$ 0.01 \\ 
8 &  0.36 $\pm$ 0.02 & 0.35  $\pm$ 0.01 \\ 
\end{tabular}
\label{table:sigmavt}
\end{table}

In addition to a well defined critical point, a true continuous phase
transition is characterized by the fact that there aren't any
characteristic length scales in the system, i.e., fluctuations take
place at {\it all} length scales. Obviously this is not true for
finite size systems -- where possibly all length scales upto the {\it
  system size}, but no larger can be present. This system-size
dependent {\it cut-off} explains why we see finite-size deviation from
the true (infinite-size) values. There is a systematic dependence of
these deviations with the system sizes studied.  By examining these
systematic dependences on the scaling exponents, we try to extrapolate
to the infinite-size value of the exponents. The fact that a
system-size independent behavior is possible over a range of system
sizes (e.g., the collapse of several system sizes onto a single curve)
is the crucial indication that finite-size scaling approach is
successful, which in turn is an indication of a phase transition.
Hence the claim that \threshvolt can be viewed as the critical point
of a phase transition.

\section{2D Arrays at Threshold}{\label{sec:paths}}

\begin{figure}
  \subfigure[a]{\epsfig{figure= \fig 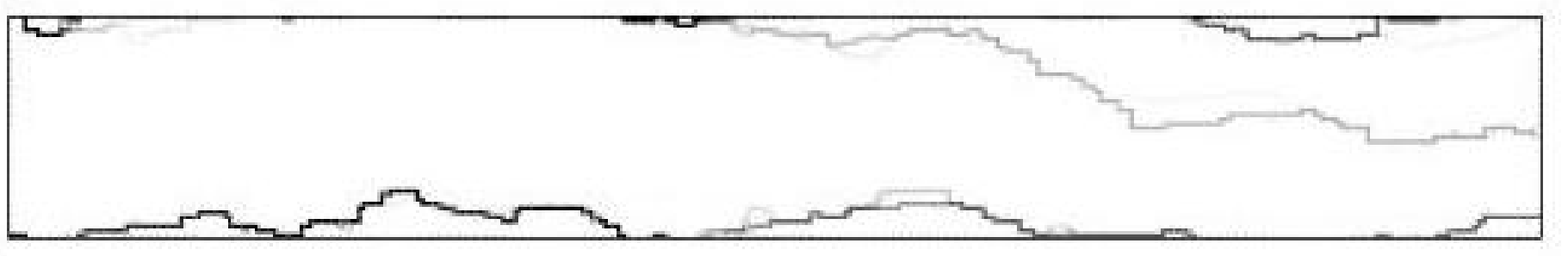, scale=0.4} \label{a}}
  \subfigure[b]{\epsfig{figure= \fig 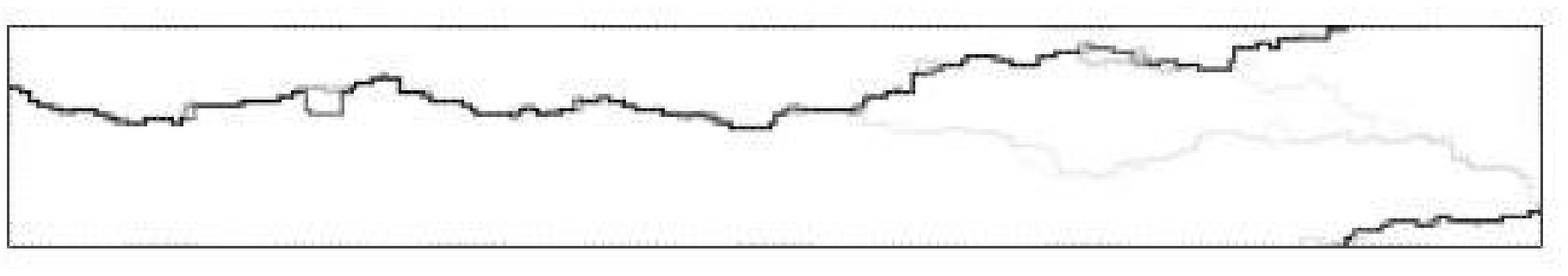, scale=0.4} \label{b}}\\
  \subfigure[c]{\epsfig{figure= \fig 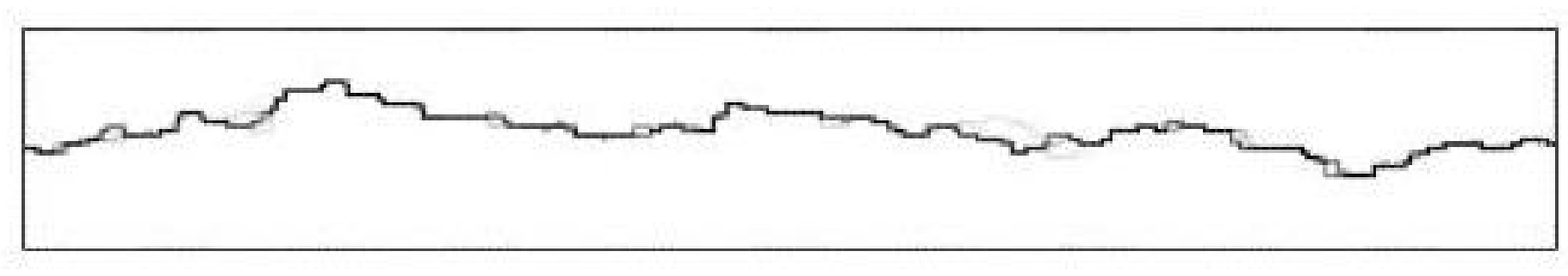, scale=0.4} \label{c}}
  \caption{A few ground state path samples illustrating the merges and splits
    along a single path.  Some splits are ``effective'' in that paths
    do not recombine, for example, the middle split in figure a. Some
    paths have multiple splits but none go onto persist till the end
    (figure c). Then there are paths that split early on and then wrap
    around -- due to periodic boundary conditions, and merge with the
    original path (figure b). This results in a mouth width equal to
    the system width. The gray-scale is an encoding for the current
    density at a given dot.}\label{sample_gpaths}
\end{figure}

$V_T$ is defined as the lowest lead voltage at which there exists at
least one dot in the column adjacent to the emitter lead, onto which
electrons can tunnel and subsequently find a way onto the collector
lead.  Having established the existence of a threshold voltage in the
previous section, our aim in this section is to understand the current
conduction in the array at {\it exactly} the threshold voltage.  A few
samples of the first conducting path
at \threshvolt 
are shown in Fig.~\ref{sample_gpaths}, from which it can be seen that
there are frequent splittings and possible recombination of paths,
leading to an overall complicated geometry and topology.  We
investigate the structure and the current carrying capacity of the
ground state paths. We will find immediate application of our
understanding in the next section, when we investigate the I-V
characteristics of 2D arrays.  In the next subsection we present the
details of how the first conducting path at \threshvolt is determined.
We then present results on the transverse deviations (meandering) of
the path -- characterized by a wandering exponent ($\zeta$) --
followed by a discussion of the main structural features of paths.
Finally, we discuss the current density profiles at the end-points and
establish a connection between structure and current density profiles
and compare the ground state path for UC arrays and DC arrays.

\subsection{Computing the ground state path} 
\label{sec:gory_details}

To describe the ground state path, in addition to determining the
location of where electrons flow, we are interested in determining the
relative probability of an electron tunneling through a given site. We
will use relative probabilities as measured by current densities ($j$)
[to be defined in Eqn.~(\ref{eq:curr_dncty})] and not macroscopic
currents (I).  We use a transfer-matrix approach to determine the
relative probability of electrons tunneling through dots. This
involves computing the probability of being in state {\it i}, using
known probabilities of being in all possible previous states {\it j}
and the transition probability of going from the states {\it j} to
state {\it i}.  Due to the numerical uniqueness of the random
potential at each site, there is in practice only a single dot onto
which electrons can tunnel from the emitter lead at $V_T$. We assign
this dot, which is at the same potential as the emitter lead, a
current density of 1.0, as all current flowing onto the array passes
through this {\it head} dot.  It can be shown that an electron cannot
tunnel onto any other dot in the left most column other than the root
of the spanning avalanche~\cite{jhathesis} -- the head dot.  Thus all
other dots in the leftmost column are assigned a probability of 0.0
(the boundary condition).

As electrons can tunnel onto a dot only from dots that have a higher
electro-static potential. Hence in order to compute the current
density of a dot (probability of an electron flowing onto a dot), the
current density of all neighboring dots which have a higher potential
should be known.  The current density of any dot i (\cdnosp), is
computed as the product of the current densities of dots in the
immediate vicinity of $i$ with the probability of tunneling from the
neighboring dot onto dot $i$, summed over all dots:
\begin{equation}
  \label{eq:curr_dncty}
  j_i = \sum_{V_{j} > V_i} j_{j} p_{j -> i},
\end{equation}
where $j_{j}$ is the current density of dot j and $p_{j->i}$ is the
probability of tunneling from a dot $j$ to $i$.  $p_{j->i}$ is
computed as,
\begin{equation}
  \label{eq:prob_tnlng}
  p_{j \rightarrow i} = {\Gamma_{j \rightarrow i}}/{\Gamma_{out}^{j}} ,
\end{equation}
where ${\Gamma_{j \rightarrow i}}/{\Gamma_{out}^{j}}$ is the ratio of
the tunneling rate from dot {\it j} onto dot {\it i} over the sum of
all outgoing rates from dot {\it j}.  As the probability of tunneling
onto a dot from a dot at lower potential is zero.  Thus starting with
the head dot with a current density of 1.0 and sorting all dots in
decreasing order of potential, the current density is computed for the
dot with next highest potential.  As the current densities and the
probabilities of tunneling are known for all dots from which electrons
can tunnel onto it, the current density for the new dot can be
determined using Eqn.(\ref{eq:curr_dncty}).

A special case of \cd is $j_L(i)$, which is defined as the current
density from the $i$th dot in the last column onto the collector lead.
It is useful to note that the sum of the $j's$ along any column can be
greater than 1.0 (e.g.  when there is intra-column hopping), but the
sum of all current densities between adjacent two columns must be
equal to 1.0 (essentially a current continuity equation). Thus the sum
of all $j_L(i)$ will be 1.0.

There isn't a simple connection between the current densities computed
using our approach and the actual macroscopic currents that can be
carried by a path. The current densities approach taken here, provides
information on the relative proportion of the current that would
tunnel off the dots onto the collector lead, i.e., be carried along
different paths, but says nothing about the exact values corresponding
to a given path. It is possible for example, that at threshold, a
simple non-splitting path conduct more current than a highly complex
path with many splittings and recombinations.

\subsection{Ground state path properties} \label{sec:phys_prop}
\subsubsection{Path meandering, widths and energy fluctuations}\label{sec:paths_wander}

\begin{figure}
  \includegraphics[scale=0.60]{\fig 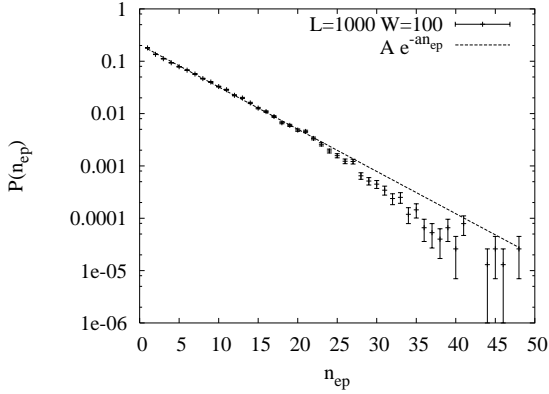}
  \caption{
    The probability distribution of the $n_{ep}$ taken
    determined using approximately 75000 disorder realizations.
    There is a good fit to an exponential line, implying 
    that there is an exponentially decreasing probability of
    a path having higher number of end-points. 
    Other system sizes have a similar distribution.\label{prob_nep.1000.100.1.0.0.0} }
\end{figure}

The number of end-points ($n_{ep}$) is defined as the number of dots
in the last column -- adjacent to the collector lead -- which have a
nonzero value of \cd (strictly speaking, a non-zero value of
$j_L(i)$).  As shown in Fig.~\ref{prob_nep.1000.100.1.0.0.0}
it becomes exponentially less probable to find paths with an an
increasing number of end-points.  It is relatively simple to implement
a tracking algorithm that by working downwards from the head node
computes the trajectory of the electron hopping and determines the
number of transverse (inter-row) deviations en-route to the end nodes.
The deviation of end-points is of interest and for the {\it i}$^{th}$
dot is given by $y_L(i)$.  The current density weighted transverse
deviation can be defined as:
\begin{equation}
  \label{eq:phii_defn}
  \phi_i = {j_L}{y_L}(i)
\end{equation}
and the current density weighted mean transverse deviation as: 
\begin{equation}
  \label{eq:Phi_defn}
  \Phi = {\sum \phi_i}
\end{equation}

\begin{figure}
  \includegraphics[scale=0.60]{\fig 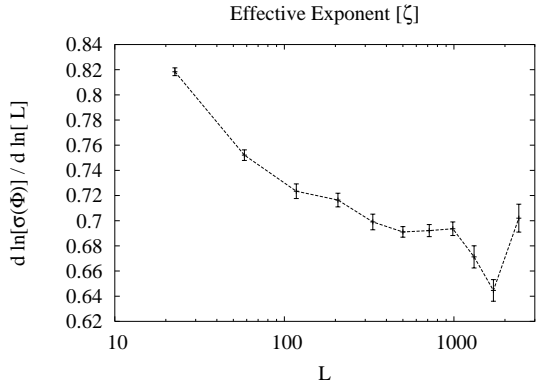}
  \caption{\label{wandering.uc} Plot of the value of
    the effective value of the fluctuation of $\Phi$ with L which
    gives the wandering exponent.  The wandering exponent approaches a
    value consistent with $\frac{2}{3}$.}
\end{figure}

As there are often more than a single end-point, thus the weighted
mean \Phy determines the location of the effective end-point of the
path and thereby the deviation of the path from the head node.  The
value of \Phy averaged over many samples will be zero, as there is an
equal probability of the effective end-point being on either side. A
look at the values of \Phy over many disorder realizations, reveals
essentially a Gaussian distribution about the mean value and the
standard deviation of the distribution grows as L$^{\zeta}$, where
$\zeta$ is the wandering exponent and is found to have a value of
$\frac{2}{3}$. This sets the scale for typical sample-to-sample
fluctuations of the effective end-point.  Fig.~\ref{wandering.uc}
plots the local value of the exponents, which in general is a useful
way to get a handle on the finite-size dependencies of the exponents.
\footnote{It can be argued that if we are simulating systems of sizes
  L $\times $ \ltth, then the wandering exponent can be no larger than
  \ftth. To establish that $\zeta$ is not system width limited, we
  simulated systems~\cite{jhathesis} with multiple basins, i.e., with
  widths greater than \ltth.  We find from a plot of the effective
  local exponent values for $\zeta$, that the value of $\zeta$ is
  still consistent with \ftth}

\begin{figure}
  \subfigure[]{\epsfig{figure= \fig 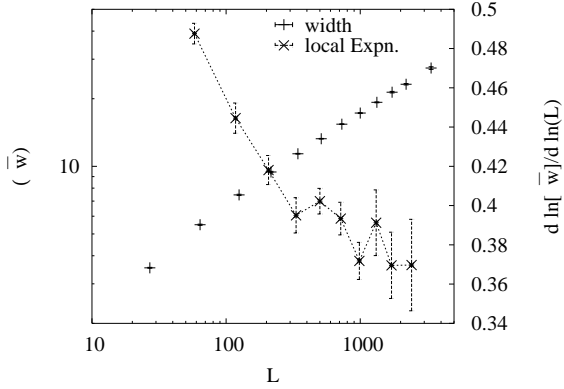, scale=0.6}\label{mouth_width_uc}}
  \subfigure[]{\epsfig{figure= \fig 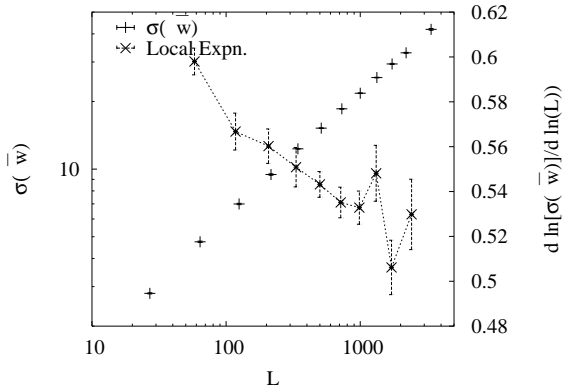, scale=0.6}\label{mouth_width_flct.uc}}
  \caption{The width of the mouth of the path is defined as the $[max(y_L(i))
    - min(y_L(i))]$.  In Fig.~\ref{troubledfig}(a) the left y-axis
    gives a plot of the width as a function of system size; using the
    right y-axis gives the local slope of the width with system size.
    As can be seen the value of the exponent settles to around 0.37
    $\pm 0.03$.  Fig.\ref{mouth_width_flct.uc} shows the results for
    the fluctuation of the width of the mouth of the path.  The growth
    in the fluctuations of the width are consistent with \lotwnosp.
    The left-hand y-axis represents the sample-to-sample fluctuations
    of the width, whereas the right-hand y-axis gives the value of the
    local slope.}\label{troubledfig}
\end{figure}

Computing the transverse deviation associated with dots helps
calculate the width of the ``mouth'' of the path in the last column,
which is defined as the difference of the transverse coordinates of
the extreme end-points.  It is of interest to understand how the width
of the mouth grows with system size. The wandering exponent provides
insight into the typical fluctuations of the location of current
density weighted end-point but does the width of the mouth grow with
the same exponent?  The mean value of mouth-widths for UC arrays is
shown in Fig.~\ref{troubledfig}(a). The growth in the width is
consistent with \loth -- definitely different from \ltthnosp.  This is
indicative of a situation where the location of the effective
end-point fluctuates as \ltth but the distribution of the extreme
end-points around the effective end-point gets ``smaller'' relative to
the effective end-point fluctuations. The increase in the mean width
of the mouth tells us that paths that require the same potential
difference as the ground-state path to within O(1), are to be found
upto \loth around the effective end-point.  The fluctuations in the
width of the mouth is plotted in Fig.~\ref{troubledfig}(b). The
fact that the fluctuation in the width scales as \lotwnosp, indicates
that the extremities of the mouth are determined by randomly juggling
end points.  We've discussed the fluctuations in the threshold voltage
in the section~\ref{sec:revisited}; we remind the reader,
that as shown in Fig.~\ref{eff_exp.threshfluct.uc} and
Fig.~\ref{eff_exp.threshfluct.dc}, the fluctuations in the threshold
voltage scales as  L$^{\frac{1}{3}}$.

As a quick consistency check that the wandering exponent is not
dependent on the width, we compared the wandering exponent for systems
with $N_b = 4$ (L$ \times N_b$L$^{2/3}$) with those of systems with
$N_b$ = 1.  Although there are significant differences at smaller
system sizes, for larger system sizes the boundary effects become less
significant. For systems with $N_b=$4, the convergence to the
$L^{2/3}$ is sooner than for $N_b=$1, indicative that boundary effects
dominate at small system lengths.

\subsection{Path geometry and topology: Gap sizes, merge lengths and 
  typical splitting distances}{\label{sec:paths_structure}}

So far our understanding of the structure of the ground state path is
that there are possibly many branchpoints leading to multiple
end-points. The current density weighted transverse deviation leads to
an effective-end-point, with sample-to-sample fluctuations of \ltth
and mouth-width which scales as \lothnosp. As a consequence of the many
interspersed end-points between the extremal end-points of the mouth,
the mouth has a fine structure not accessible by studying only the
transverse fluctuations of the effective end-point and widths.  We
would like to understand the details of fine structure of the mouth,
viz.,  to understand if any two physically contiguous end-points are
part of the same path segment or if they belong to two different path
segments.  In general, if two end-points belong to different segments,
typically how far back did those segments separate?  Answers to these
questions, will help understand several important length scales
characterizing paths at threshold. It will also provide
additional metrics to compare the ground-state paths of systems with
different disorders (UC, DC and RT).

\begin{figure}
  \subfigure[]{\epsfig{figure= \fig 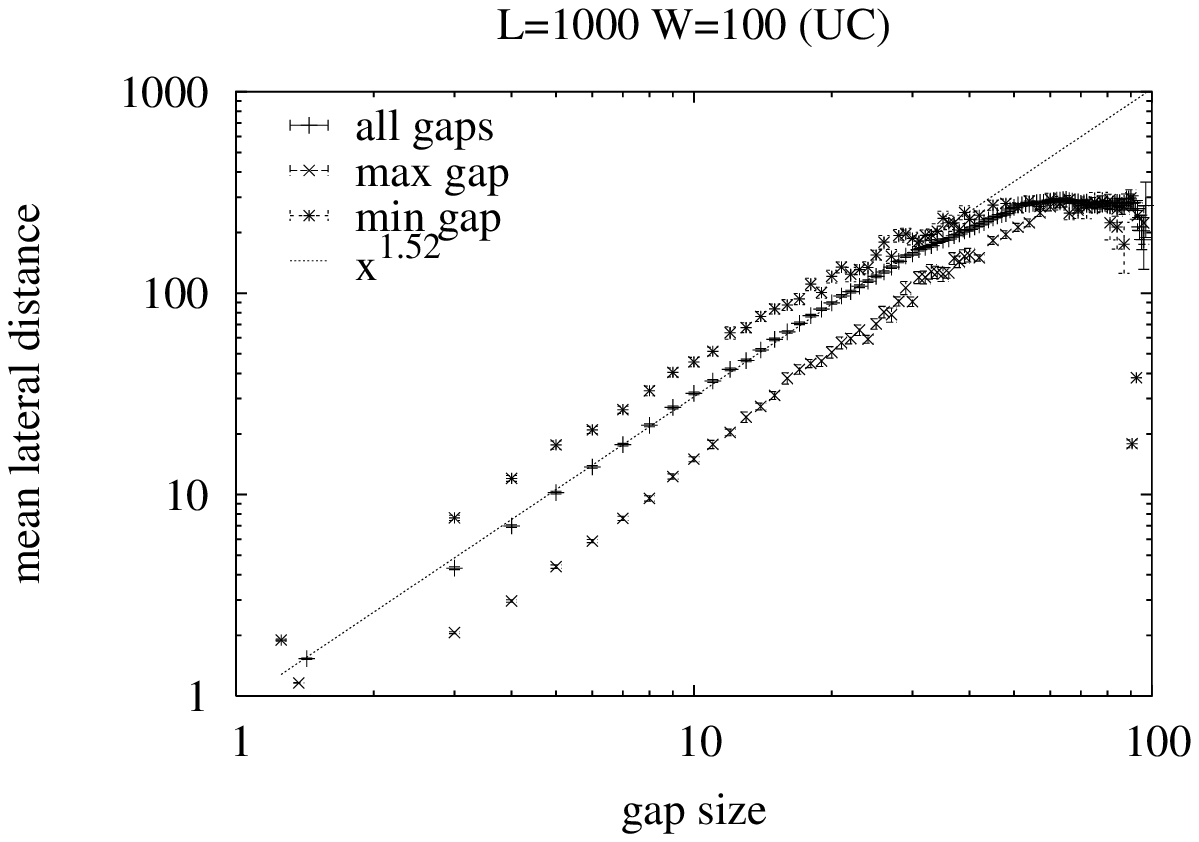, scale=0.60} 
    \label{comp_alldelta_l.1.0.0.0}}
  \caption{ 
    Fig.~\ref{comp_alldelta_l.1.0.0.0} is a plot showing the mean
    lateral length for all gaps of size $\Delta$.  The mean lateral
    length of a gap of size $\Delta$, can be thought of as the typical
    distance at which the path to the head-node from two end-points a
    distance $\Delta$ apart will have merged. The fit indicates that
    the lateral length scales as $3/2$ of the gap size. The plot also
    shows a comparison of the mean lateral lengths for the maximum
    and minimum gaps for an effective branchpoint.  }
\end{figure}

In order to compute the size of gaps separating the end-points and to
compute how frequently and over what length scales paths split we need
to look beyond the transverse deviations of each point.  We need to
track the complete trajectory electrons may take before tunneling onto
the end-points.  This involves knowing the location of all
branchpoints along the path , where a branchpoint maybe defined as a
location at which a split occurs, i.e., where there is more than one
neighboring dot onto which electrons can tunnel.  There are many
splits along the path; the majority of the splits along the path do
not survive and merge a short distance after splitting.  In the
thermodynamic limit, not all branchpoints are of interest, but only
those branchpoints that go on to produce end-points -- do not merge
after splitting.  We find the last possible location that is common to
the trajectories associated with the two end-points of interest.  This
point is referred to as the effective branchpoint corresponding to the
two end-points.  Having thus determined the location of the effective
branchpoints for all pairs we can compute the position at which paths
to the end-points last overlap.  Equivalently, this location can be
used to establish a lateral distance from the collector lead that
paths from end-points a transverse distance $\Delta y_L$ apart will
most likely merge by.

We then compute the mean value of the lateral length of gaps of size
$\Delta$.  Given that the wandering exponent has a value of
$\frac{2}{3}$, one would expect that the paths to two end-points
separated by $\Delta {y_L}$ would be typically joined upto a distance
${\Delta {y_L}}^{1/{\zeta}}$ from the collector leads.  This is
analogous to the typical separation of the optimal paths to two ends
of a DPRM that are $\Delta y_{L}$ apart, viz., ${\Delta
  {y_L}}^{1/{\zeta}}$).  As shown in
Fig.~\ref{comp_alldelta_l.1.0.0.0} our findings are in good agreement
with this expectation; the mean lateral length for all the $\frac{n
  (n-1)}{2}$ gaps for the $n$ end-points is used in the fit. The fact
that the path structure of QDA is similar to the scale-invariant tree
structure of DPRM, is further indication that ground state paths of
QDA are in the same universality class of DPRM. The significance of
this conclusion will be explored later.  Also plotted are the mean
lateral lengths of the maximal and minimal sized gaps of an effective
branchpoint, which scale like $\frac{5}{3}$ and $\frac{4}{3}$
respectively with the transverse size of the gaps.

\begin{figure}
  \subfigure[]{\epsfig{figure= \fig
      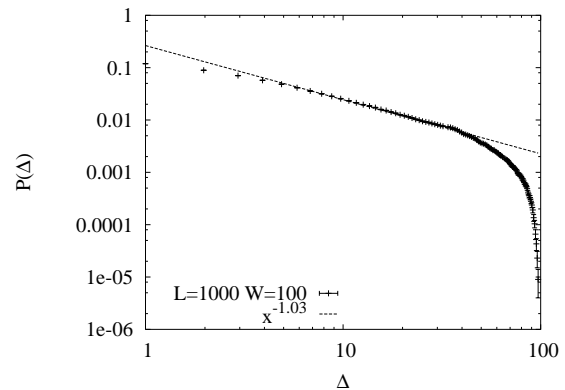, scale=0.6}
    \label{fit.histo.delta.1000.100.1.0.0.0}
  } \subfigure[]{\epsfig{figure= \fig
      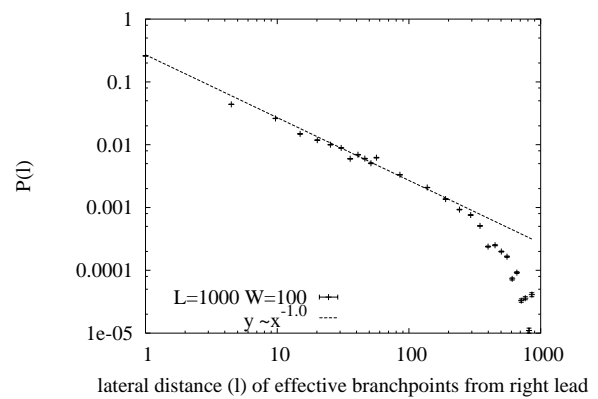, scale=0.6}
    \label{fit.histo.l.1000.100.1.0.0.0}
  }
  \caption{
    Fits to the probability distributions of the gap sizes
    (Fig.~\ref{fit.histo}(a)) and the lateral length of the gaps
    (Fig.~\ref{fit.histo}(b)) for a single system size.  Although not
    shown, there is not a system size dependence at which probability
    decreases (though there is a system size dependent cut-off).  Fits
    indicate the probability of occurrence of a gap of size $\Delta$
    decreases as ${\Delta}^{-1}$.  A similar dependence is observed
    for $l$, which is consistent with expected probability
    distribution for $l$ computed using a variable transformation from
    $\Delta$ to $l$ (where $\Delta \sim l^{2/3}$).  Given the
    distribution of $l$, the chance that paths that split, will
    survive as independent paths all the way to the end gets smaller
    the earlier they split.}\label{fit.histo}
\end{figure}

The gap between two end-points is defined as the number of the
intermediate dots separating them.  Thus for two physically adjacent
end-points (irrespective of whether they belong to the same path
segment or not), the gap is defined to be of zero size.  We computed
the effective branchpoints for all pairs of end-points (there are
$\frac{n (n-1)}{2}$ pairs for n end-points) and computed the lateral
and transverse sizes of the gaps.  The results of the probability
distribution for a fixed system size are shown in
Fig.~\ref{fit.histo.delta.1000.100.1.0.0.0} and
Fig.~\ref{fit.histo.l.1000.100.1.0.0.0}.  The probability
distributions represent the simple fact that large gaps resulting from
earlier permanent splittings of the path at the threshold become less
probable.  As the path at threshold and the end-points have become
reachable within the last $O(1)$ increase in potential, all path
segments at threshold must be equal to each other to within voltage of
$O(1)$.  Consequently they will overlap for the most part.  We have
seen that the sample to sample fluctuation in the threshold voltages
scales as \loth which should set the scale for the typical difference
between non-overlapping paths, thus at threshold we expect
typically $O(1) \over L^{\frac{1}{3}}$ 
fraction of   paths segments to not  overlap.
Consequently if for a given system size,  we were to  plot the mean of
all lateral sizes of the gaps as a fraction of the system length, we'd
expect to see a $L^{-\frac{1}{3}}$ dependence. 

By studying the distribution of the location of the effective
branchpoints, and found that it became increasingly improbable that
they would be located closer to the emitter lead~\cite{jhathesis}.  It
is also useful to compute the number of effective branchpoints (depth)
encountered on a path to an end-point and how the depth varies for the
different end-points in a given sample. This tells us if the paths to
the end-points are typically similar, as well as permitting us to
determine a correlation between physical proximity of end-points and
difference in depths.  Thus we determine the average number of
end-points for a given mean depth of end-points.
\begin{figure}
  \subfigure[]{\epsfig{figure= \fig 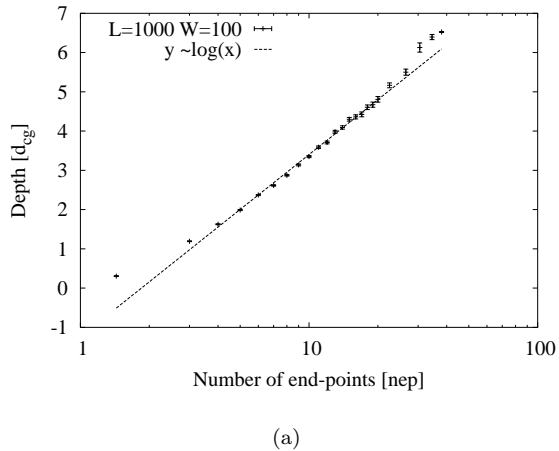, scale=0.60} 
    \label{nep_dcg}}
  \caption{
    In Fig.~\ref{nep_dcg_all} the  mean value   of  the $d_{cg}$  of   the
    end-points is plotted  as a function of   $n_{ep}$. As the  number of
    end-points  increases the  mean depth of   the end-points increases
    logarithmically. This   is indicative of  an  essentially balanced
    tree. For   a  perfectly balanced  tree  the  number  of end-points
    increases  exponentially with   the depth, whilst  for an hierarchically
    split tree (in the limit of a perfectly  unbalanced tree) the mean
    value  of  the   depth increases  linearly   with the    number of
    end-points.}\label{nep_dcg_all}
\end{figure}
If the tree structure was perfectly random the number of end-points
would grow like the square root of the depth.  On the other hand for
an essentially unbalanced trees (where all the splittings take place
on the path to one particular end-point), the number of end-points,
$n_{ep}$ grows linearly with the mean depth. For an essentially
balanced tree each path splits essentially with equal probability in
which case the number of end-points grows as some number to the power
of the mean depth (for a perfectly balanced tree this would be the
number would be 2).  As shown in Fig.~\ref{nep_dcg_all}, we find that the
mean depth grows logarithmically with the number of end-points,
characteristic of an essentially balanced tree.

\begin{figure}
  \includegraphics[scale=0.60]{\fig 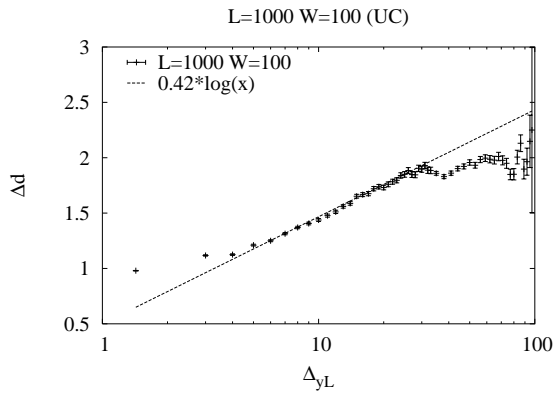}
  \caption{\label{DyL_cgdepth.1000.100.1.0.0.0}
    $\Delta y_L$  is the transverse separation of  two end-points on the
    path and $\Delta d_{cg}$ is the difference in the $d_{cg}$ between the
    nodes.  Given  the logarithmic increase in   the  value of 
    $\Delta d_{cg}$   with transverse  separation,  the   paths  to the  nodes
    typically have a net difference of one additional branchpoint with
    every scale of two increase in the  separation between nodes. 
  }
\end{figure}

Finally we'd like to determine if the trees (representing the
ground-state paths) are spatially homogenous and if the path segments
from the different end-points to the head node, are essentially
similar in the number of effective branchpoints encountered.  To do
so, we investigate the difference in the depth (on the path to the
end-point), represented as $\Delta d_{cg}$, of end-points separated by
a transverse distance $\Delta y_L$.  As shown in
Fig.~\ref{DyL_cgdepth.1000.100.1.0.0.0}, the difference in depths
increases logarithmically with transverse separation of end-points.
Given the logarithmic dependence of $\Delta d_{cg}$ on transverse
separation of end-points, the paths to the nodes typically have a net
difference of one additional branchpoint with every scale of two
increase in the separation between nodes.  

It is fair to assume that path segments that are not immediate
descendents of the same parent are essentially independent; every pair
of end-points with a $\Delta d_{cg} > 1$ can be considered independent
and thus the plot in Fig.~\ref{DyL_cgdepth.1000.100.1.0.0.0} provides
a measure of the number of independent paths segments (channels) that
reach the collector lead. This could possibly be experimentally
verified by studying the spectrum of the discrete current at the
collector lead.  Using this definition of 'independent' paths, we find
that the number of independent channels increases logarithmically in
the transverse direction upto the width of the mouth.

\subsection{Current densities}{\label{sec:currentdensities}

We have  found  that  the structure and  the   topology of  the  first
conducting   path   to    have   several  interesting  features    and
characteristic  length scales.  An  important question is  what is the
profile  of   the current  densities at the end-points?   
Also, what does the fluctuations of the current densities within the
mouth tell us about the overall structure of the paths?  

To address what the current density values for a pair of
end-points tell us, we take $j^>$ and $j^<$ as the values of the
larger and smaller current density (for the two end-points in
consideration) respectively, and as a measure of the difference in the
number of splittings encountered for two end-points define $\Delta
n_s$ as:
\begin{equation}
  \label{eq:delta_ns_defn}
  \Delta n_s = \log({j^{>}\over j^{<}})
\end{equation}

\begin{figure}
  \subfigure[]{\epsfig{figure=\fig
      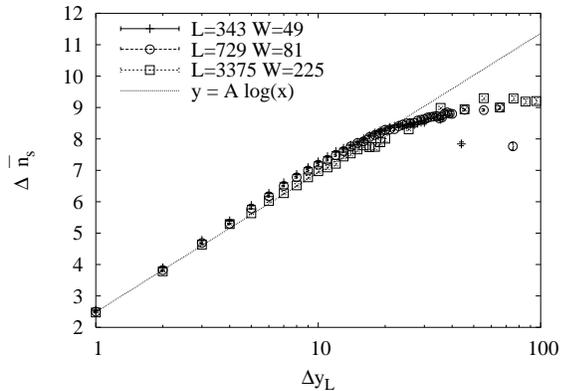, scale=0.60}
    \label{delta_yL_delta_ns.uc}} \goodgap
  \caption{
    $\Delta n_s$ is a measure of the difference in the number of
    splits in the paths traversed to get to two end-points.  In
    Fig.~\ref{delta_yL_delta_ns.uc} the dependence of the mean value
    of $\Delta n_s$ on the separation $\Delta y_L$ is plotted.  As can
    be seen initially there is a logarithmic dependence of $\Delta
    n_s$ on $\Delta y_L$ (the fit shown here is for the largest system
    size) before gradually crossing over to a $\Delta y_L$ independent
    value.  Data for widely differing system sizes (from L=343 to
    L=3375).  To within errors, the plateau value appears to be
    independent of the system size.}\label{delta_yL_delta_ns.uc.all}
\end{figure}
In Fig.~\ref{delta_yL_delta_ns.uc.all}, we plot the value of $\Delta n_s$
as the transverse separation between end-points increases.  The best
fit to the largest system size considered (L=3375, W=225) is
consistent with a logarithmic dependence over the range 1 to about 20.
From the logarithmic increase in $\Delta n_s$ with transverse distance
($\Delta y_L$), it follows that $j^{>} = \Delta y_L j^{<}$, i.e., with
increasing distance between the two points considered, the larger
current density ($j^{>}$) tends to get larger relative to the smaller
($j^{<}$) -- increasing linearly in $\Delta y_L$.  We know from insight
gained from the structure of the paths, that as the transverse
distance separating two end-points increases, the paths taken to the
two end-points separate earlier. If the paths to the end-points after
separation typically undergo the same number of current splittings,
then on average there would not be any variation in $\Delta n_s$ with
distance; but given the slow but definite distance dependence, it is
consistent to conclude that one path undergoes more current splits
than the other, and that for end-points separated by a greater
transverse distance, the correlation in current densities will be less
than for those end-points which have greater overlap in their
paths.

It is useful to point out the similarity between the logarithmic
dependence of $\Delta n_s$ on $\Delta y_L$ as in
Fig.~\ref{delta_yL_delta_ns.uc} and the logarithmic dependence between
the $d_{cg}$ on $\Delta y_L$ as shown
Fig.~\ref{DyL_cgdepth.1000.100.1.0.0.0}.  In general, given the
similarity in the properties of the current densities at the
end-points and the structure of the path, it appears to be the case
that the effective branchpoints not only determine the structure of
the paths but also play a role in determining the current profiles of
the end-points.

\subsection{QDAs with capacitance disorder} {\label{sec:path_ucrd}}
   
In this subsection, we study the properties of the first path for DC
arrays and begin by investigating the transverse deviations of the
path and the structure of the mouth for ground-state paths.
\begin{figure}
  \subfigure[]{\epsfig{figure= \fig 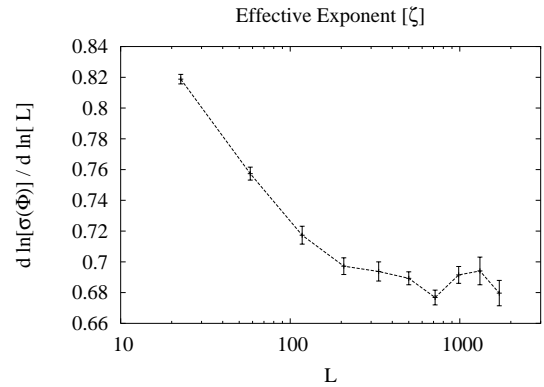,
      scale=0.6} \label{wandering.dc}} 
  \subfigure[]{\epsfig{figure=
      \fig 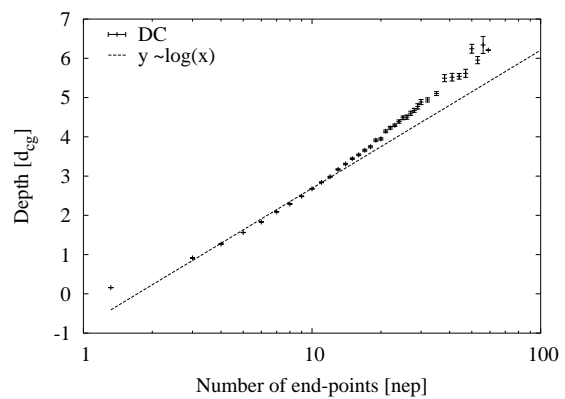, scale=0.6}
    \label{UC_DC.nep_dcg}}
  \caption{The values of the local slopes for the wandering exponent 
    computed for arrays with disordered capacitance are plotted in
    Fig.~\ref{wandering.dc}.  From Fig.~\ref{UC_DC.nep_dcg} $d_{cg}$
    increases logarithmically with $n_{ep}$. Both are essentially
    indistinguishable from UC arrays.}\label{path_dc_all}
\end{figure}
As shown in Fig.~\ref{path_dc_all}(a), the wandering exponent gradually
approaches the value of $\zeta=$ \ftthwsp for larger systems, which is
similar to UC arrays.  In Fig.~\ref{UC_DC.nep_dcg}, the relationship
between $d_{cg}$ and $n_{ep}$ is shown to be logarithmic (recall that
the probability of occurrence decreases exponentially as $n_{ep}$
increases).

In  Figs.~\ref{UC_DC.delta} we plot the distribution of  the gaps and the mean
lateral    distance of splitting for gaps of size  $\Delta$.  Both the distribution of  the gaps sizes (and  thereby
lateral size of the gaps) and the mean  lateral distance dependence on
gap sizes are similar to UC arrays.

\begin{figure}
  \subfigure[]{\epsfig{figure=   \fig   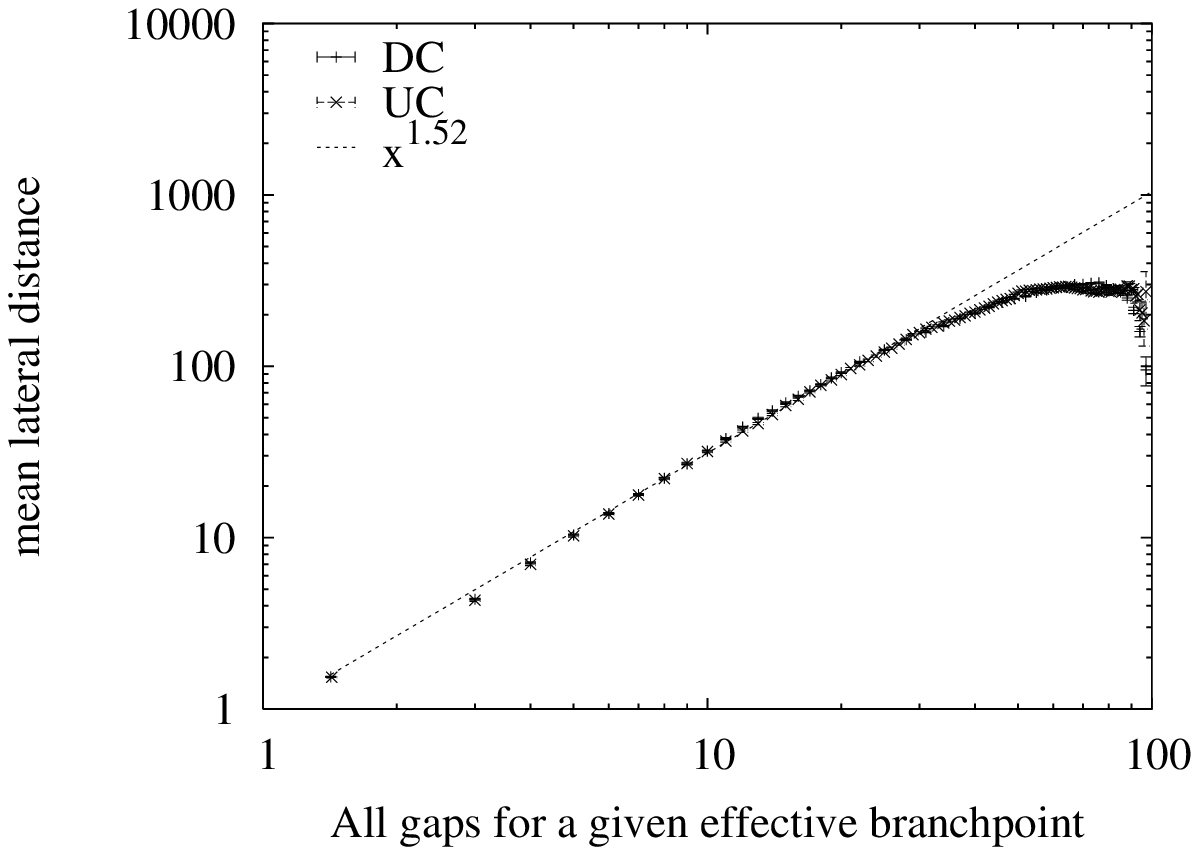,
      scale=0.60}   \label{UC_DC.alldelta_l}} 
  \subfigure[]{\epsfig{figure= \fig 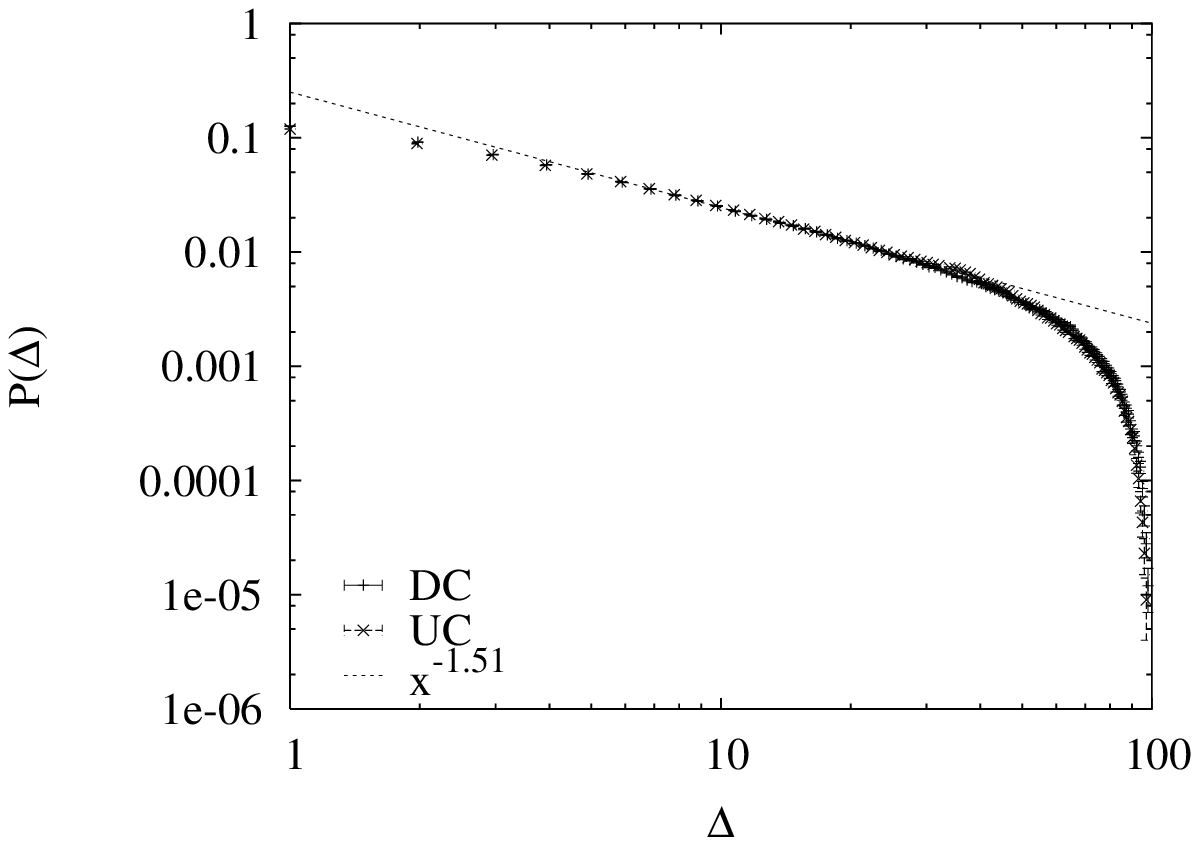, 
      scale=0.6}   \label{UC_DC.alldelta}}
  \caption{Comparing the gap distribution and spacing for UC and DC in
    Fig.~\ref{UC_DC.alldelta_l} and Fig.~\ref{UC_DC.delta}(b),
    indicates that they are essentially indistinguishable.}
\label{UC_DC.delta}
\end{figure}

In addition to the structure of the path, current flow properties are
also indistinguishable to UC systems as shown by the sample averaged
fluctuations of the current-density weighted transverse
locations~\cite{jhathesis}.
From the data as presented in this section, ground state paths are
effectively indistinguishable from the UC. It is highly unlikely that
any further investigations will indicate any significant differences
between the ground path structures for the UC and DC systems.

DPRM is controlled by a zero fixed point thus the ground state (lowest
energy) strongly determines the properties of the system. Given the
fact that the ground-state conducting path is in the same universality
class as the DPRM, one would expect excited conducting paths -- those
with energy higher than the ground state at higher voltages as well as
ground-states at non-zero temperatures -- to be strongly influenced by
the structure and energetics of the first conducting path. Thus the
connection between QDA and DPRM in addition to providing an idea of the
structure of the ground state path at zero temperature, indicates that
an extension of the approach used here to study the ground state path
might possibly be used in determining sensitivity to boundary
conditions and temperature changes of the ground state paths.  The
latter is of significant practical importance.  Given the putative
similarity between QDA and DPRM we can use results obtained in the
DPRM case to predict a temperature sensitivity: namely that the ground
state configuration is sensitive to temperature changes and will most
likely rearrange. As to whether this is sufficient to change any
scaling properties will require explicit numerical and analytic work.

\section{Conduction in 2D Arrays \label{sec:2d_dynamics}}

In the previous sections, we saw how the threshold voltage can be
viewed as the critical point of a continuous phase transition and
explored the associated critical phenomenon {\it at and below the
  critical point}.  This sets the stage to address the next, and
arguably most important question in our investigation of disordered
QDA -- the nature of the critical phenomenon for voltages {\it above
  the critical point}.  Based based on the strength of the driving
force ($V$) relative to the strength of internal interactions and
disorder strength, roughly three distinct regimes can be identified.
The first regime can be thought of as when the scales of disorder,
interaction and driving force are all similar.  In this regime the
role of disorder is generally crucial and the interactions between the
many degrees-of-freedom result in strong deviations from a mean-field
behavior.  This regime typically occurs when $V$ is very close to the
threshold voltage.  A second regime lies at the other end of the
spectrum, where the driving force is extremely strong compared to the
strength of disorder and interactions between the degrees of freedom;
in this regime the disorder and interactions become irrelevant and the
system is driven into a linear response mode.  Our {\it primary} focus
will be on the investigation of the critical behavior and dynamical
response close to the transition -- corresponding to the first regime.
We will study the dynamic response by computing the I-V
characteristics for a range of different systems sizes.  Details of
theory and implementation of our numerical simulations can be found in
Ref.~[\onlinecite{jhathesis}].

The relative strength of the interactions and disorder in turn has
been used to broadly classify two widely differing types of collective
transport: weak disorder relative to the strength of interactions most
likely leads to an elastic structure without breaking up; an example
of which are CDW. In general when the disorder is strong relative to
the interactions, the elastic structure breaks and transport is far
more inhomogeneous and plastic like. Examples of transport in such a
plastic regime include, the flow of a non-wetting fluid in porous
medium~\cite{porous_fluid}, the transport of strongly pinned
two-dimensional Abrikosov flux array~\cite{plastic3}, driven
collective transport of neutral carriers in randomly varying
traps~\cite{plastic2, plastic6} and the flow of a fluid with no
elastic interactions flowing down a rough inclined
plane~\cite{plastic5} (the dirty windshield problem).  We will find
that conduction in the low $\nu$ regime is plastic-like, i.e., along
well defined narrow channels.

Recall that below $V_T$, the concept of an advancing elastic interface
-- defined as the contour of maximum advance of charge along a given
row was useful.  As a consequence of our definition, this elastic
interface is no longer well defined at driving voltages above
threshold, and thus not the interface that tears and results in
plastic flow. This leads to an interesting situation where as a
consequence of asymmetry around threshold, the variables and
description of the system on the opposite sides of the critical point
are different; consequently the same exponents are not valid both
above and below the transition point. This is unlike many continuous
phase transitions, especially equilibrium (e.g., two-dimensional Ising
magnets in the absence of an external field) but even non-equilibrium
phase transitions (e.g., CDW) where the same exponents with possibly
the same values characterize the critical regimes on either side of
the critical point.

It is instructive to review the MW scaling hypothesis originally
presented in Ref.~[\onlinecite{mw93}] to understand current flow in a
two-dimensional arrays before discussing the numerical results.
Similar to one-dimensional arrays, any current-carrying channel at a
given emitter lead voltage $V_L$, there will be $\frac{V_L - V_T}
{(\frac{e}{C_{\Sigma}})}$ extra charges on average. The locations of
these extra charges can be viewed as charge steps relative to the
threshold configuration.  Where exactly these extra charges are
located on the channel depends on the underlying disorder; typically
charge down-steps are where the tunneling rates are sufficiently
smaller than the mean tunneling rates. The location of these steps
give the {\it most likely } locations of a split in the path; and thus
can be used to define a correlation length $\xi_\parallel$, where
$\xi_\parallel = \frac{eL}{(V_L - V_T)C_{\Sigma}}$, where $L$ is the
linear dimension between the emitter and collector leads. We have seen
that the transverse deviation ($\xi_\perp$) of a path segment of
length $\xi_\parallel$ is given by $\xi_\parallel^{2/3}$.  Also $V_T
\sim L$, thus $\xi_\parallel \sim {\nu}^{-1}$ and therefore $\xi_\perp
\sim v^{-2/3}$.  $\xi_\perp$ sets the scale for separation between
channels {\it before splitting}.  The number of channels ($N_{ch}$) at
the collector lead will thus be given by $\frac{W}{\xi_\perp}$, where
W is the width of the array.  Under the assumption that each channel
reaching the collector acts as an independent one-dimensional
current-carrying chain, the current in a channel is $I_{ch} \sim \nu$.
Thus the total current carried by the array will be given as:
\begin{equation}{\label{eqn:mw_hypo}}
  I \sim {N_{ch}   \times I_{ch}}  \hspace{0.2in} \sim  {\nu}^{5/3}
\end{equation}  

It is important to discuss some of the assumptions that the MW
hypothesis depends critically on.  

Firstly, that each channel behaves like a one-dimensional array and
the current in the 1D array grows linearly with $\nu$.  Secondly, that
number of channels grows like $\sim {\nu}^{2/3}$, which in turn is
dependent upon two assumptions. The first is that the transverse
deviations grows like $l^{2/3}$, where $l$ is a linear dimension of
the path. We have extensively verified this to be true {\it at} $V_T$;
it is fair to assume that it is above $V_T$ too.  The second
assumption is that the most likely effective splits -- splits that do
not result in a recombination -- take place at the charge down-steps.
This has been harder to verify rigorously; at best we find for arrays
at $V_T$ that the sample-averaged probability of effective splits
decreased as $\sim \frac{1}{l}$.  We find that the fluctuations in the
current carrying capacity of one-dimensional arrays decreases as $\sim
{\nu}^{-1/2}$

It was originally predicted \cite{mw93} that to observe the true
exponent arrays larger than $400 \times 400$ would be required. We
will go onto show that the linear dimension required before the ``true
exponent'' might be observed is probably an order of magnitude larger
than initially estimated. A significant portion of the remainder of
this section will be devoted in support of this statement.

\subsection{Simulation results, analysis and discussion}
\label{sec:dynamics}

From our analysis of the avalanches and path meandering, we developed
an understanding that on average, a L$ \times$L$^{\frac{2}{3}}$ sized
system contains one independent basin and thus one channel.  Before we
can understand the current carrying capacity of several interacting
channels, it is important to determine how the current carrying
properties of a single channel changes with driving voltage.
Consequently, we will most often investigate the I-V curves of
asymmetrical systems of length L and width L$^{\frac{2}{3}}$.  Due to
finite size effects and computational cost, this also happens to be a
practical approach.

We are investigating the scaling properties of a hypothesized power
law between I and $\nu$, thus rather than perform a coarse grained fit
to the entire I-V and generate a single value for the scaling
exponent, we compute ``local'' values of the exponent $\beta$ as a
function of $\nu$.  We find this a more useful and meaningful
representation for determining how the scaling relation between the I
and $\nu$ changes.  The procedure although helpful, is not
sufficiently sophisticated to be a complete replacement for a rigorous
fitting, as error bounds with confidence intervals are not easily
determinable from this approach.

\begin{figure} 
\subfigure[]{\epsfig{figure= \fig 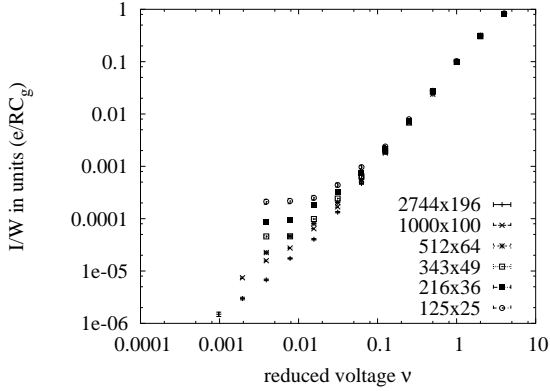,       scale=0.6}
  \label{allcurr.1.0}}   \goodgap   
\subfigure[]{\epsfig{figure= \fig 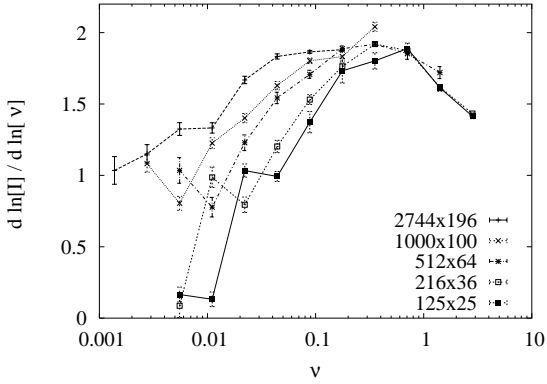, scale=0.6} 
  \label{all_run_slope.1.0}}
\caption{
  I-V curves for 2D arrays with offset charge disorder for a range of
  system sizes.  The \localexp for the I-V curves in
  Fig.~\ref{all_curr_uc}(a) is plotted in Fig.~\ref{all_curr_uc}(b).
  For a true power law scaling, the value of \localexp for different
  system sizes should overlap. As can be seen this does not happen for
  values less than $\nu = 0.1$.  Following the MW scaling hypothesis,
  we expect a plateau at values less than $\nu = 0.1$. The inability
  to see clearly a definitive plateau is primarily due to large finite
  size effects.}\label{all_curr_uc}
\end{figure}

The local exponents for UC arrays are plotted in
Fig.~\ref{all_curr_uc}(b), from which there is a clear dependence
on system size for the local exponents. There isn't a range of $\nu$,
however small, where the local exponent curves for all the different
system sizes lie on a single curve, as would be expected for a valid
scaling regime.  Thus it is difficult to claim that there is a {\it
  unique single} value of the local exponent for all sizes, even over
the smallest regime of $\nu$.  At lower values of $\nu$ the
statistical noise starts to dominate and the true value of the local
exponent becomes unclear.

The aim of rigorously verifying the MW scaling hypothesis numerically
does not appear to be easily attainable with available computational
resources at the present moment; thus it remains open, as to
whether the MW scaling hypothesis is valid.  If we assume that MW is
the correct hypothesis, we can at best determine the constraints on
system sizes and values of $\nu$ to establish a regime for the
validity of the hypothesis.  This is somewhat analogous to determining
an upper bound of the reduced variable upto which critical behavior
can be observed:

It is known that for CDW one has to be within $f \leq 0.01$
\cite{thorne_pt} (where $f = {(F- F_c)\over F_c}$) of the critical
point in order to observe associated critical phenomenon. Similarly
for high-T$_{c}$ superconductors (copper oxide) in three-dimensions,
by some estimates~\cite{goldenfeld} the critical region exists for $t=
10^{-4}$ where $t = {(T- T_c)\over T_c}$ The quoted estimates are from
analytical calculations and supported by numerical data. With the
caveat that it is much harder to estimate correctly using numerical
data alone, we hazard an estimate of the critical region for QDA
solely on numerical data. From the plot of the local slopes for the
largest two-dimensional QDA simulated ($L=2744$ and $W=196$ in
Fig.~\ref{all_curr_uc}(b)) and the plot of \localexp for largest
one-dimensional systems (L$=2000$) a similar upper bound would be
somewhere between $\nu = 0.01 - 0.1$ for L$\times$L$^{\frac{2}{3}}$
arrays.  As can be seen from Fig.~\ref{all_curr_uc}(b), there are
strong indications of a plateau in \localexpnosp, albeit over a small
region -- for values of $\nu$ around 0.1 -- and only for the largest
arrays.  In addition, from the very brief flattening out of \localexp
for $1000 \times 100$ arrays around $\nu =0.1$ before dipping, it is
conceivable that for values of $\nu < 0.1$, the ``true exponent''
value lie somewhere in between 1.5-2.0; this is consistent with the
hypothesized value of \localexp $=5/3$.
If at all, this will be the critical region and the likely value of
the critical exponent.

It is clear that simulations of even larger system sizes will be
required to observe a plateau for at least a decade in $\nu$ -- which
is the really the minimum range over which a power law should be
observed before definitive claims of scaling are valid.  As mentioned
simulations of systems large than $2744 \times 196$ are currently
computationally not feasible.

For values of $\nu > 0.1$ the values of \localexp are influenced by a
crossover to a peak value of approximately $2.0$, before being driven
into the linear regime.  This bump in the values of the \localexp
corresponds to a regime outside of the putative MW regime, when new
splits in the current carrying channels are taking place at all length
scales and thus there are rapidly increasing new outlets giving rise
to the value of $2.0$ for the \localexpnosp.  New channels open, but
are not all independent; for L$ \times$L$^{\frac{2}{3}}$ arrays these
newly opened channels will typically merge with the ground state path.
The effective value of \localexp at 2.0 appears to be a coincidence, a
malicious one for several experiments seem to encounter this value too.
The effective exponent value at a given $\nu$ is sensitive to the
ratio of the length to widths, albeit in a complex fashion.

As a consequence of finite-sizes, a crossover region over which the
effective exponent is different from the ``true exponent'' arises.
The crossover region gets larger for smaller system sizes.
\begin{figure}
  \subfigure[]{\epsfig{figure=\fig 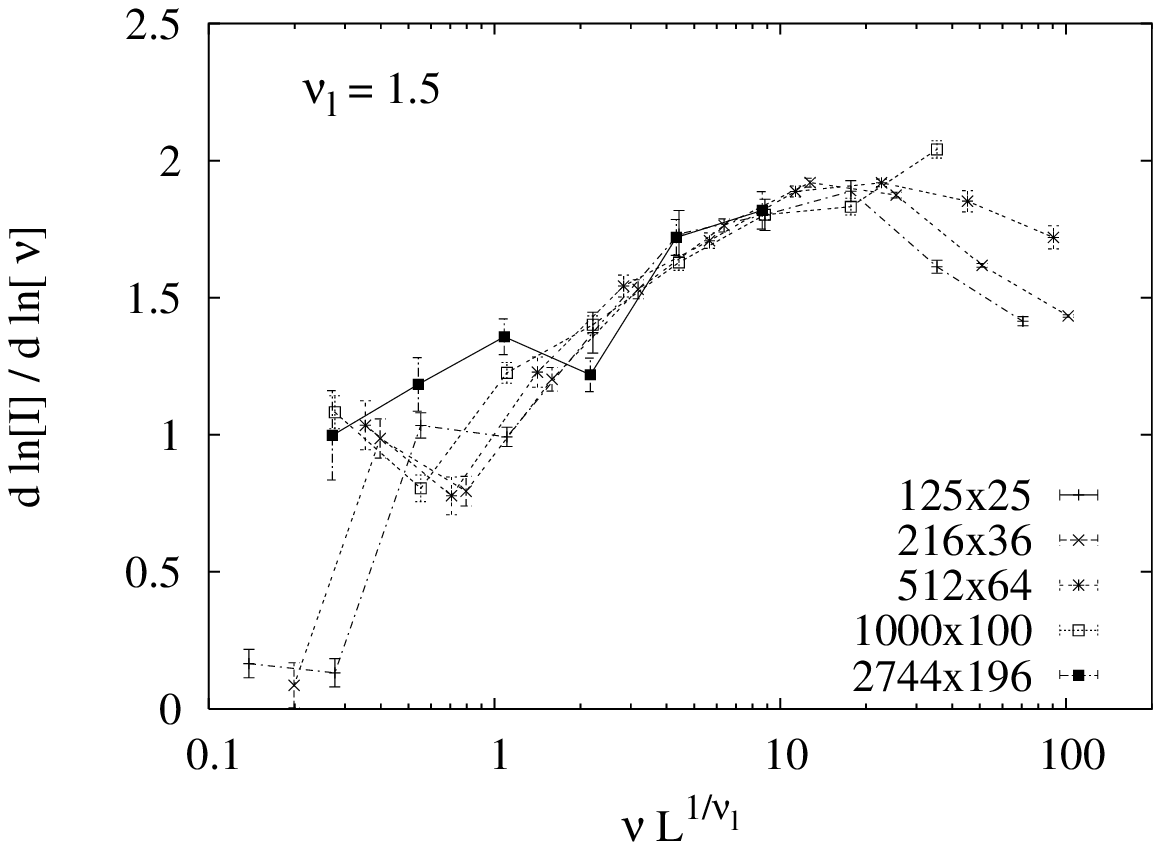,
      scale=0.6}
    \label{scale_local_slope-8.1.0}}\goodgap
  \subfigure[]{\epsfig{figure=\fig 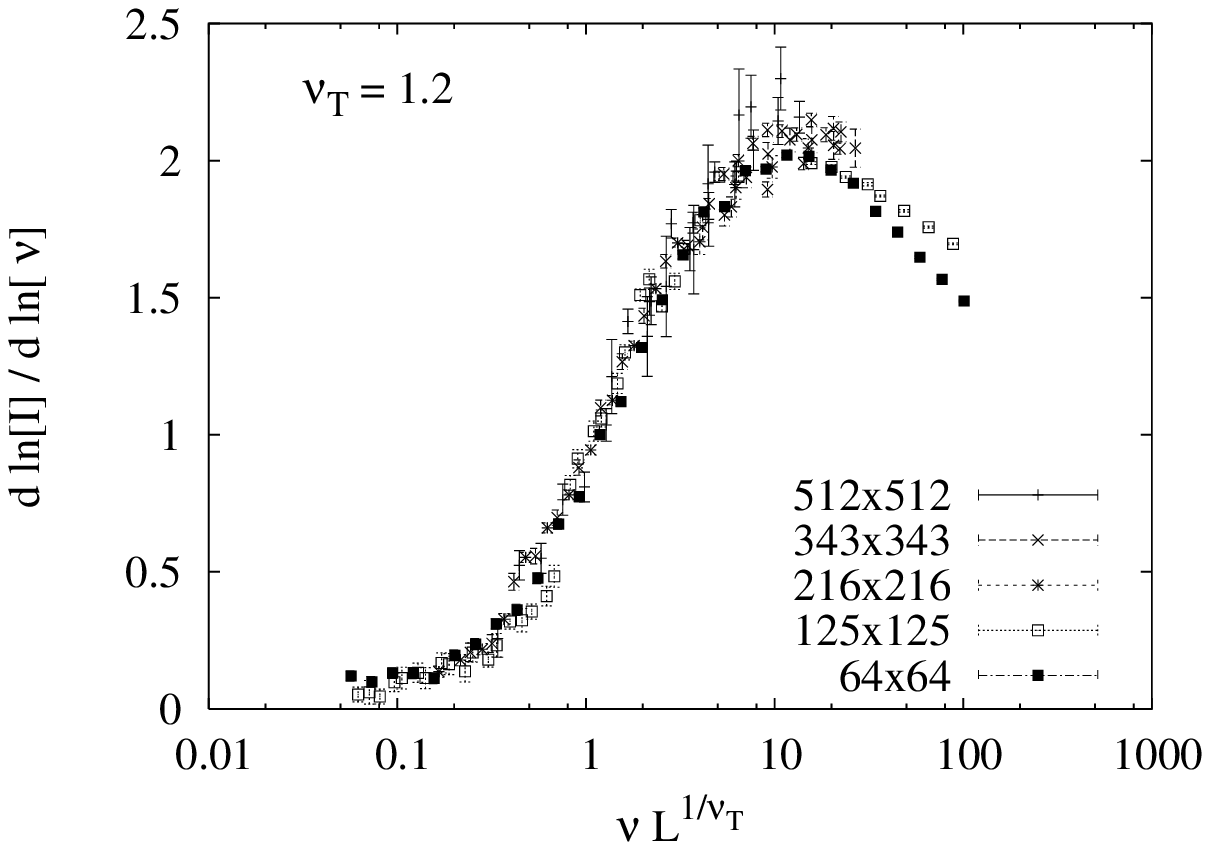,
      scale=0.6}
    \label{symm_scale_local_slope-8.1.0}} \\
  \caption{
    The collapse of the local slopes for uniform capacitance systems
    is plotted in Fig.~\ref{scale_local_slope_uc_all}(a).  Collapse of
    the local slopes for symmetric systems with uniform gate
    capacitance is plotted in Fig.~\ref{scale_local_slope_uc_all}(b).
    Note that the value of ${\nu}_T$ that gives good data collapse is
    different from the value of ${\nu}_T$ that gives similar collapse
    for systems of size L x
    L$^{2/3}$.}\label{scale_local_slope_uc_all}
\end{figure}
Somewhat analogous to the finite-size scaling exponent $\nu_{T}$
characterizing fluctuations in the threshold voltage, we attempt to
define a finite-size exponent $\nu_{l}$, which helps characterize this
crossover region over which values of the \localexp for arrays
L$\times$ L$^{\frac{2}{3}}$ deviate from the true exponent.  From the
plot in Fig.~\ref{scale_local_slope-8.1.0} we find that the best
estimate is given by $\nu_{l} = 1.5$.  Although the quality of the
collapse is by no means satisfactory, a couple of trends are
noticeable: there appears to be a a region over which the \localexp
appears to lie on a single curve (roughly over $\nu L^{1/{\nu_{l}}}$
values from 1 to 10) and one notices that larger systems appear to
stay on the collapsed curve upto smaller values of $\nu
L^{1/{\nu_{l}}}$.  It is interesting to note that {\it it appears}
that $\nu_{l}$ is similar to $\nu_{T}$, which if true would imply the
existence of single finite-size length scale. Also, for small systems
at large $\nu$, there are strong signs from the values of \localexp
that the transition towards linear behavior has begun.

There are at least a couple of factors possibly preventing the
observance of a true scaling region.  Firstly, there are very strong
finite size effects.  Systems of a sufficient size are required before
the putative scaling behavior can be discerned.  For example, 1D
channel sizes need to be long before the linear dependence of I on
$\nu$ can be observed. We estimate from our analysis in
Sec(\ref{sec:onedim})that they should be at least longer than 1000
dots. So although using brute force computational power we have
reduced significantly the statistical noise for smaller systems (e.g.,
$343 \times 49$, $216 \times 36$) , these systems sizes are
insufficient to actually observe the putative scaling and we observe
an effective exponent not in agreement with the theoretically expected
scaling values. Secondly, statistical noise needs to be reduced
significantly further for larger systems.  The reduction of
statistical noise for large system sizes, especially at lower $\nu$,
is strongly dependent on the cost of determining {\it correctly} the
value of the current. We will elaborate on this further.

Additional complications arise from the fact that there are large
fluctuations and large timescales associated with channel formation.
It is the complexity associated with both determining correctly the
channel structure as well as the converged current value that makes
the exact and proper simulation of electron flow in arrays such a
difficult task.  The two issues are in someways aspects of the same
problem -- the timescales required to form a steady state current
pattern are long and broadly distributed between samples.  This
phenomenon is common to several other dynamical systems involving
collective transport and disorder~\cite{plastic3, plastic2}.  The
timescale required for current patterns to reach a steady state
appears to be different from the timescale required for the current
values to reach steady state.  For a particular sample considered, the
difference in current after the last channel formed was only $2 \%$,
but while investigating systems of the same size and at similar
$\nu$'s (to within a factor of 2), we noticed that when channels
formed, there were concomitant changes in the current by more than $20
\%$.  This wide variation is part of the problem -- for it is
difficult to estimate how much, a well formed channel will contribute
to the overall current. Any adaptive algorithm based upon channel
formation and activity isn't easy. As L gets larger the problem gets
more acute.  A somewhat similar problem, is the long time scale
required to form a channel, even if there is just a single channel
involved in conduction.

As a consequence of the above features, it can be difficult to
determine the current value correctly.  At best, we can strive to
minimize the probability of getting an incorrect current value. As
with other simulation schemas, it soon becomes a problem of optimizing
a finite amount of resources -- the reduction of statistical noise has
to be traded off with systematic errors.  Naively one would expect
that the channels that conduct most of the current would form early
on, and thus with simulations of {\it sufficient duration} the major
current carrying channels will have reached a steady-state, both in
terms of current carried and formation.  This is not necessarily the
case and even if it were, given the broad range of times for this to
happen between samples of a given size, it would require setting all
simulation runs to be sufficiently long to accommodate the longest
time to steady-state.  In addition to being difficult to estimate {\it
  a priori}, it would be computationally no more efficient than using
an algorithm that determines dynamically whether the current channels
have reached a steady-state.

After accounting for initial transient effects, we set a bin size to
be 10000 and compute the current in the first two bins, based upon
which we use a convergence criteria (to be described later) to
determine if the current has reached a steady state. If the current
hasn't converged, the bin-sizes are doubled, i.e., the number of
electrons that we wait for to tunnel off are doubled, after which
similar checks to determine the steady state is done on the next two
bins.  This process disregards the history and values of previous
bins.  One of the reasons this is done, is because it can take very
long for the steady-state distribution to be free of initial
transients and biases.  It is difficult to use two successive bins
{\it from the same initial configuration for convergence} to determine
in a definitive way whether we have reached a steady state.  Our
approach to correctly determining steady-state current, is to use two
different initial conditions and to simulate until they each reach
values that: (i) individually converge, and (ii) converge with respect
to each other.  (This is somewhat analogous to the situation for
simulations of glassy systems where at least two different starting
configurations are adopted as a measure to check against getting stuck
in a local minima while exploring state space).  We refer to the two
starting states as the ``hot'' and ``cold'' configurations
respectively.  The classification of hot and cold states reflects the
fact that the cold initial configuration has been prepared by the
addition of electrons so as to have a smooth spatial gradient of
electron potential from the emitter lead to the collector lead for the
given value of $\nu$, while the hot initial configuration has a smooth
spatial gradient of electron potential but corresponding some value
greater $\nu'$ than the required value of $\nu$.
  
The rate at which the local values of current change can be very
different for hot and cold states; it is also typically very different
for different samples. It is possible that a simple measure of
convergence like setting an acceptable upper limit on the percent
difference between the values of local current in two bins before
considering the current to have converged, mistake the slow change to
be an incorrect convergence. Any measure of convergence whether hybrid
or for a single state should take into account the fluctuations in the
value of the local current.  Our method for determining convergence
can be summarized as follows: After reaching threshold, we initialize
two different states -- hot and cold.  We start by simulating the hot
configuration until it converges after which we switch to the cold
configuration.  We check if hot and cold states satisfy the
convergence criteria so as to distinguish it from the convergence test
of two successive bins from the same starting convergence (single
convergence).  Every time the cold configuration is checked for single
convergence.  If the cold configuration is singly converged, but the
hybrid convergence criteria isn't satisfied then we switch to the
previously saved hot configuration.  In general, we check for hybrid
convergence every time a single convergence check is performed and
toggle between hot and cold states every time either one of them
satisfies single convergence but the test for hybrid convergence isn't
satisfied.  This is an attempt to keep the dynamical evolution of the
hot and cold simulations somewhat in phase. By ensuring that the
individual configurations have separately converged -- once the
test of hybrid convergence is passed, it is fair to assume that we
have determined the steady-state current value and pattern.  For the
same net computational resource, the hybrid convergence method
provides higher quality data~\cite{jhathesis}.

Related to the ongoing analysis of statistical versus systematic
errors, we present some final remarks about simulations of larger
system versus more disorder averaging: just because the ability to
simulate larger systems may exist, does not make it necessarily an
efficient use of computational resources. Bigger may not always be
better -- for it maybe possible to get more accurate results by
simulating larger number of samples (disorder realizations) of smaller
system sizes than smaller number of samples of larger system sizes.

A system is said to be self-averaging \cite{binder+landau} for a
variable A, if the error in $n$ statistically independent measurements
of A ($\Delta$A) tends to go to zero as $L \rightarrow \infty$, i.e.,
\begin{equation}
  \Delta A(n,L) = \sqrt{(<A^2> - {<A>}^2)/n}, \hspace{0.2in} n \gg 1
\end{equation}
If it goes to an L-independent value the system lacks self-averaging.
We computed $\Delta V_T(n, L)$ for $n=10000$ samples and we find that
one-dimensional arrays are {\it strongly} self-averaging, i.e.,
$\Delta V_T(n, L) \sim {(nL^d)}^{-1/2}$, where $d=1$. Two-dimensional
arrays are weakly self-averaging: in this case $\Delta V_T(n, L) \sim
{(nL^{-2/3})}$.  Generally if a system is self-averaging than
simulations of larger system sizes is not counter-productive.
Recapitulate that from plots of \localexp in one-dimensions, we saw
that linear chains of lengths greater than a 1000 are required to
observe linear scaling; this in turn in a way sets a lower bound on
the sizes of two-dimensional arrays.

\subsection{Disordered capacitance}

\begin{figure}
  \subfigure[]{\epsfig{figure=\fig 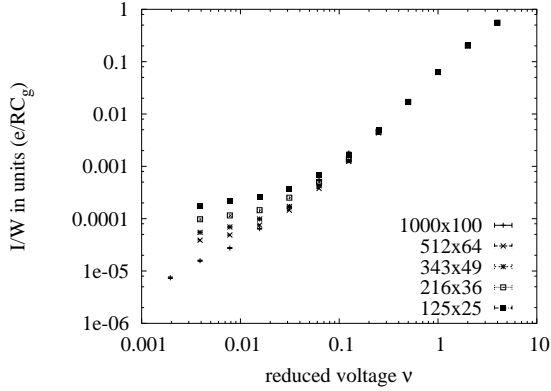,
      scale=0.6}\label{all_curr.2.0}}
  \subfigure[]{\epsfig{figure=\fig 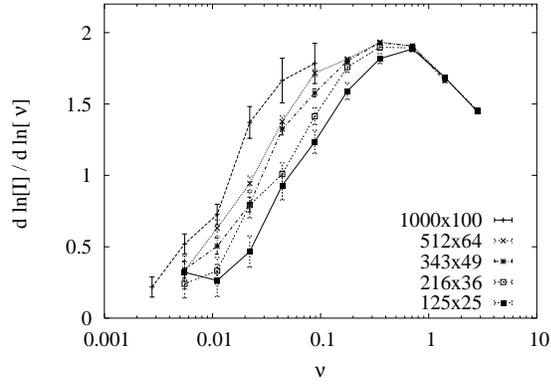,
      scale=0.6}\label{all_run_slope.2.0}}
\caption{I-V curves for systems with disordered capacitance are plotted in
  Fig.~\ref{all_curr_dc}(a).  Local slopes of the I-V curves are
  plotted in Fig.~\ref{all_curr_dc}(b).}\label{all_curr_dc}
\end{figure}

We plot the I-V curves and the $\beta_{local}$ for DC arrays in
Fig.~\ref{all_curr_dc}(a). As can be seen in Fig.~\ref{all_curr_dc}(b)
there does not exist a regime where a definite single value of the
exponent describes the scaling of I with $\nu$ for all system sizes.
Unlike UC systems, we have not simulated arrays of size 2744 $\times$
196 but only upto 1000$\times$100 -- which goes to substantiate the
dependence of the putative scaling exponent on system sizes.

\section{Other Results}\label{sec:or}

\begin{figure}
  \center \includegraphics[scale=0.6]{\fig
    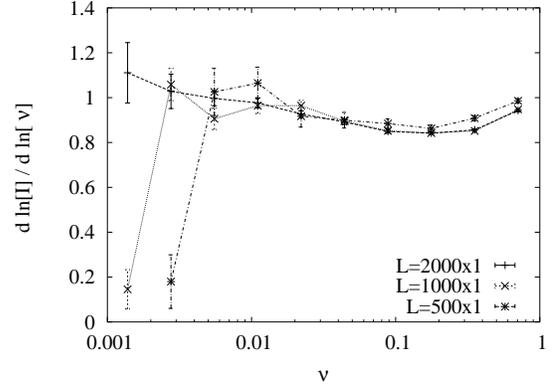}
  \caption{
   \label{all_run_curr.1.0.1.0.1D} The \localexp for
   1D arrays with offset charge disorder and tunneling resistance
   disorder. The effective exponents are quantitatively similar to 1D
   arrays without tunneling disorder.  For larger system sizes the
   value of \localexp approaches 1 at smaller $\nu$.  Although the
   slow points are a consequence of a combination of tunneling
   resistance fluctuations and small voltage differences, the basic
   mechanism of overcoming slow points with increasing voltage remains
   and thus the linear dependence on increasing $\nu$.}
\end{figure}

We briefly discuss results for for one-dimensional arrays with
tunneling disorder. Along with the understanding of paths at threshold
from section~\ref{sec:paths}, we can use it to infer some basic
features of the ground-state path of RD arrays, even though we have
not explicitly simulated 2D systems with tunneling disorder.  The
\localexp for 1D arrays with tunneling disorder is shown in
Fig.~\ref{all_run_curr.1.0.1.0.1D}.  The \localexp are qualitatively
and even quantitatively similar to 1D arrays without tunneling
disorder. We note that the value of \localexp approaches 1 at smaller
$\nu$ for larger system sizes. The system sizes and values of $\nu$ at
which they approach 1 are essentially similar to 1D arrays without
tunneling disorder.  So although the slow points now are a consequence
of a combination of tunneling resistance fluctuations and local
minimums in the potential gradient (rate differences), the basic
mechanism as outlined earlier for 1D arrays of overcoming slow points
with increasing voltage remains valid and thus the linear dependence
on increasing $\nu$. This combined with results of the transverse
deviation of paths at threshold, where we found that $\zeta$ scales as
L$^{\frac{2}{3}}$, irrespective of the type of disorder, indicates
that {\it a priori} there is no reason to expect that splittings will
occur any differently (pre-factors may change) and thus the
probability is very small that I-V scaling on introducing tunneling
disorder will be any different in the thermodynamic limit.

\begin{figure}
  \subfigure[]{\epsfig{figure=\fig 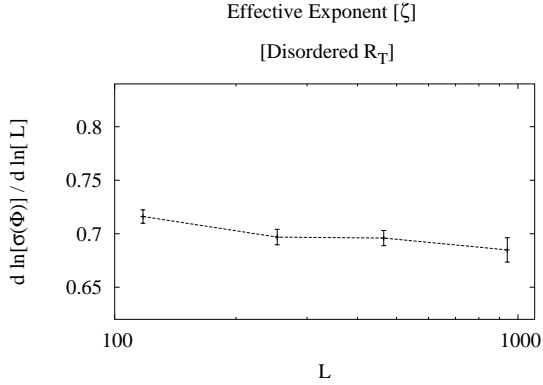,
      scale=0.60} \label{wandering.rd} }
  \subfigure[]{\epsfig{figure=\fig 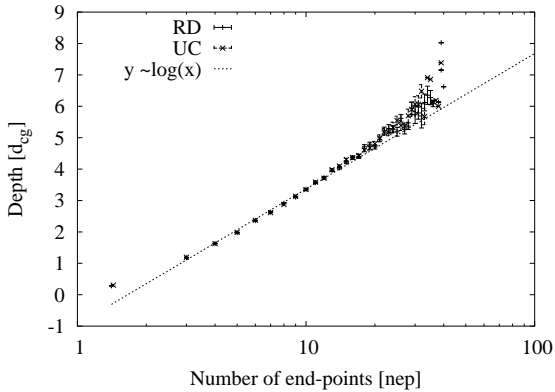, scale=0.60}
    \label{UC_RD.nep_dcg} } 
  \caption{Fig.~\ref{wandering.rd} plots the dependence of the standard 
    deviation of $\Phi$ with L for arrays with both offset charge
    disorder and random tunneling resistances is plotted.  Note that
    the in spite of the introduction of resistance disorder the
    scaling exponent of the transverse meanderings is not different
    from the meandering of the first path for UC arrays.
    Fig.~\ref{UC_RD.nep_dcg} shows a comparison of the relationship
    between $n_{ep}$ and $d_{cg}$ for UC and RD arrays. A logarithmic
    dependence ($d_{cg} \approx \log(n_{ep})$) holds for both.}\label{otherresults.1}
\end{figure}

Similar to the comparisons of $\zeta$ between DC and UC arrays, we
compute the wandering exponent for RD arrays.  We find that the value
of the wandering exponent $\zeta$, as shown in
Fig.~\ref{otherresults.1}(a), approaches the value of $\frac{2}{3}$ as the
size of systems simulated gets larger.  Also, as shown in
Fig.~\ref{otherresults.1}(b), the structural properties of the
ground-state path as measured by the relationship between the depth
and the number of end-points is similar to that of UC.  The
probability distribution of gaps and the mean lateral length of
separation for gaps of size $\Delta$ are plotted in
Fig.~\ref{UC_RD.alldelta} and Fig.~\ref{UC_RD.alldelta_l}
respectively.  Differences with UC if any are not significant.  From
the comparisons between UC and RD arrays as well as UC and DC arrays,
it can be confidently said that the main features of the ground-state
path -- meandering, structure and geometry -- are invariant to the
type of underlying disorder.

\begin{figure}
  \subfigure[]{\epsfig{figure= \fig 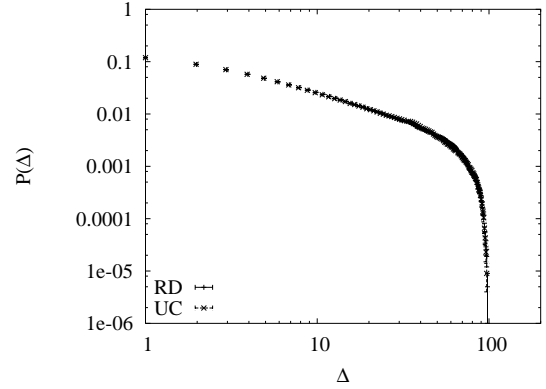,
      scale=0.6}
    \label{UC_RD.alldelta} }
  \subfigure[]{\epsfig{figure= \fig 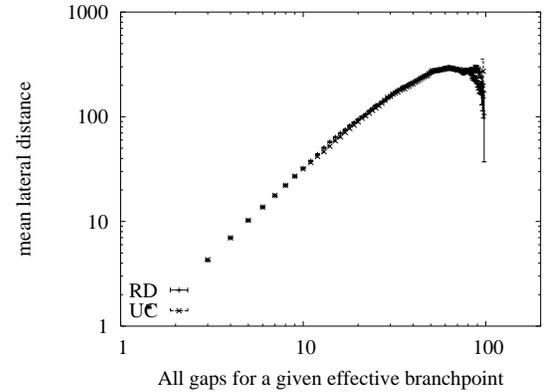,
      scale=0.6} \label{UC_RD.alldelta_l} }
  \caption{Plots comparing the probability distribution of gap sizes
    and the dependence of mean lateral length with gap sizes for UC
    and RD arrays. The mean lateral length scales as
    ${\Delta}^{\frac{3}{2}}$ for RD arrays as shown in
    Fig.~\ref{UC_RD.alldelta_l} whereas, Fig.~\ref{UC_RD.alldelta}
    compares the probability distribution of the gap sizes for UC and
    RD arrays.}
\end{figure}

There have been suggestions based on experiments
Ref.~[\onlinecite{jaeger01}] that the presence of the tunneling
disorder could lead to greater transverse fluctuations because of the
introduction of additional possible bottlenecks arising due to the
large fluctuations in the tunneling resistances. Based upon numerical
simulations, we do not notice any changes from the properties of UC
arrays in the transverse meandering or the structure of the paths.
\begin{figure}
  \subfigure[]{\epsfig{figure= \fig
     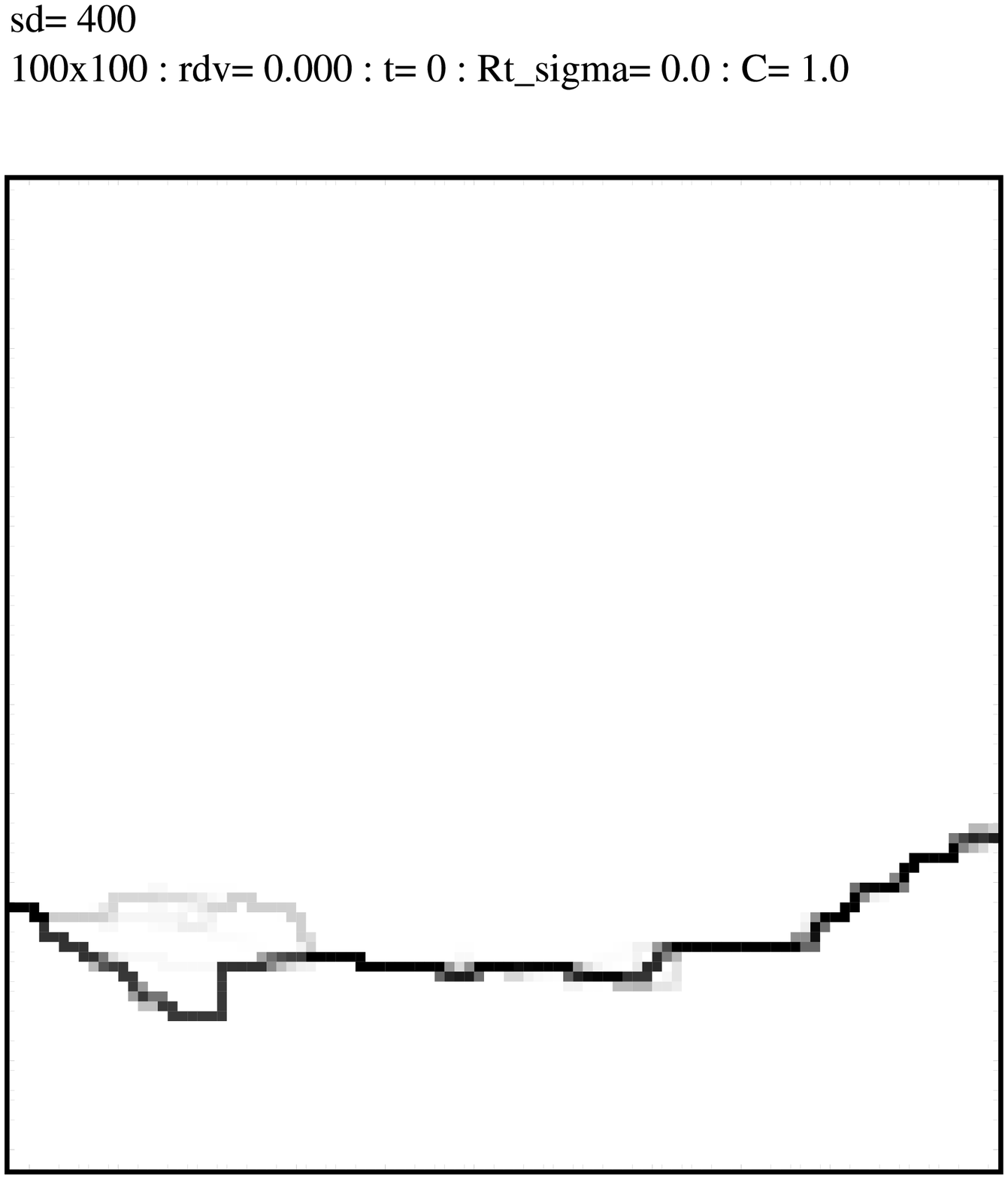,scale=0.25}
    \label{compare_gs_path.uc}} \subfigure[]{\epsfig{figure= \fig
      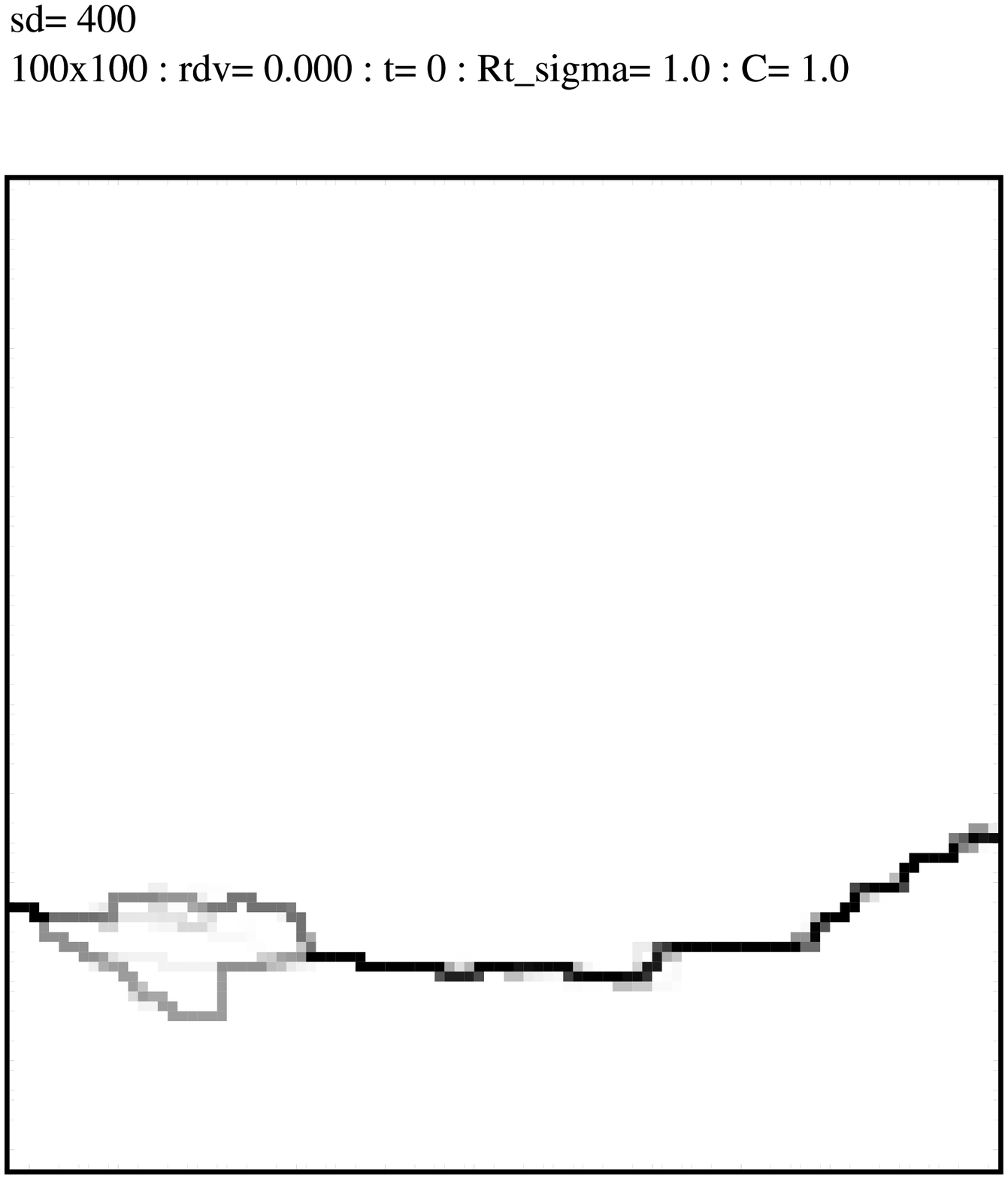,scale=0.25}
    \label{compare_gs_path.rt}}
\caption{Comparing the ground state path at $\nu = 0.0$. 
  Fig.~\ref{compare_gs_path}(b) is an array with exactly similar
  charge disorder as the array in Fig.~\ref{compare_gs_path}(a), but
  with tunneling resistance disorder included.  Although the current
  densities at various locations are different, the overall structure
  is similar.}\label{compare_gs_path}
\end{figure}
To validate further the claim we carried out the following numerical
experiment: we first computed the path in a few UC arrays.  Keeping
everything else the same, we introduced disorder in the tunneling
resistances in the otherwise similar arrays and computed the paths.
For the four different arrays we experimented with, we did not find
any significant changes in the structure of path (although actual
values of the current densities will be different).
Fig.~\ref{compare_gs_path} shows the results for one of them.
Although not conclusive, this is indicative that resistance disorder
at most changes the current density distribution for a given sample
and that change is indistinguishable when averaged over many samples.
It is important to mention that the ground state paths in
Fig.~\ref{compare_gs_path} and similar experiments, were not computed
using the transfer-matrix approach, but was dynamically determined at
$\nu = 0.0$. It is possible, however, that due to greater dynamical
freedom in selecting current flow paths at higher values of $\nu$,
there still be differences in the properties of current carrying
paths.

\section{Summary and Conclusions}\label{sec:sac}

Using computer simulations we can easily control the presence of
different disorder types and thus discern the individual and
collective effects. In doing so, we find that the presence of
background charge disorder is the dominant type of disorder, and
although there are some minor changes for arrays with variable
capacitance and tunneling disorder, the main scaling arguments and
exponents characterizing the arrays at \threshvolt and in the
conducting regime close to \threshvolt remain unchanged.  A study of
the interface properties in section \ref{sec:subthresh} indicated that
the ground-state path for two-dimensional QDA should belong to the
same universality class as the DPRM.  By looking at the structure and
the transverse deviations of the ground-state path we were able to
establish the connection conclusively.  We also saw in
sections~\ref{sec:onedim} and~\ref{sec:paths} that the introduction of
disordered \csigma does not change the current-scaling exponents for
one-dimensional arrays nor of the ground-state paths. From
section~\ref{sec:2d_dynamics}, it appears that the scaling exponent
$\zeta$ for 2D arrays does not depend upon the types of disorder
simulated either.

The dominance of charge disorder is probably due to the fact that the
disorder energy scale is set by the presence of the background charge
impurities, is the crucial energy scale of the system.  This in part
is due to the fact that the fluctuation between the charging energy of
dots as a consequence of the particular parameter values we choose
($C_{\Sigma}^{max}$ = 2.0), is less than the fluctuation in
electrostatic energy due to offset charges being chosen randomly
between [0,1[. The presence of tunneling disorder does not change the
energetics of the arrays, i.e., $V_T$ and fluctuations in $V_T$. It is
important to remark, however, that if the offset-charge impurities
were disregarded and only a non-uniform \csigma considered, the arrays
would still exhibit a threshold voltage, separating the insulating and
conducting phases and most properties would still be similar to the
situation where there was only offset-charge impurities.  If the
energy fluctuations due to of non-uniform \csigmanosp, was greater
than the background charges the claim would be that the dominant form
of disorder was the \csigma {\it although the properties of the array
  would be essentially insensitive to which was the dominant disorder}.


From our discussion in section~[\ref{sec:2d_dynamics}], we make the
important conclusion that it is most likely that the MW hypothesis is
correct and valid for disordered QDA irrespective of the actual
relative strengths of the disorder. It is important to appreciate
that one needs {\it to get sufficiently close to threshold to observe
  the scaling and that too only for large systems}.

From our experience, a naive approach to determining the steady state
current consistently underestimated the current values, which tends to
get more acute at lower values of $\nu$. As a consequence, a higher
putative value of $\beta$ would be observed.  It also follows that
there is a need for careful simulations, for with slightly less
diligence it would have been tempting to predict an exponent range of
2.0-2.25.  This is intertwined with the issue of high computational
cost, which arises from a combination of the need to compute the
converged current accurately for a {\it single} sample and the need to
simulate large systems as a consequence of strong finite-size effects.
It is not easy to formulate an elegant algorithmic solution to this
problem.  Although, parallelization is a well defined and often used
approach to reduce time-to-solution of a problem, our problem does not
appear to be a suitable candidate, for as mentioned, one of the
primary bottlenecks in our simulations is the extremely long times
required to reach a steady-state configuration. It is
physically-meaningless to run a simulation at a time $t_2$ without
state information at time $t_1$, where $t_2$ is a time later than time
$t_1$; thus there is a fundamental limitation on speed-up that can be
achieved via parallelization.  But parallelization possibly along the
lines of Ref.~[\onlinecite{korniss00}] may be a possible route forward.

We have focused on QDA in the extreme limit where the screening-length
is less than the spacing between dots. The opposite regime of
essentially infinite screening-length has been well studied, both
numerically and theoretically for non-disordered arrays
\cite{bakhalov89} and recently for arrays with a random background
potential~\cite{likharev_prb03} -- although neither of these studies,
nor others that we are aware of, use the {\it statistical physics}
approach that we have used.  Surprisingly, there has been little
activity in the regime representing the middle ground, viz., a
screening length from a few upto a dozen dot spacings.  With a
screening-length more than a single dot spacing, the on-site
interaction model that we have used in this work is not valid and
computational approaches will require fundamental reworking.
Ironically this regime is important (and interesting), as most
nanoparticle arrays as a consequence of the absence of an underlying
gate {\it most probably} have an electrostatic screening-length of a
few dot spacings.

In summary, we have investigated the effect of disorder on the
transport of electrons in arrays of mesoscopic sized metallic islands,
at, below and above a critical voltage $V_T$.  In contrast to
experiments, using computer simulations we can easily control the
effects of different disorder types.  We find that the presence of
background charge disorder is the dominant type of disorder and
although there are some minor changes with the addition of variable
capacitance and tunneling disorder, the main scaling arguments and
exponents characterizing the arrays at threshold and in the conducting
regime remain unchanged. Our numerical results indicate a value for
the exponent $\beta$ to be in the range 1.5-2.0.

\begin{acknowledgments}
  One of us (SJ) would like to thank Dave McNamara for helpful
  discussions during the initial stages of the work.  This work was
  supported in part by the National Science Foundation
DMR-0109164. 

\end{acknowledgments}
\bibliographystyle{apsrev} 
\bibliography{qdots}
\end{document}